\documentclass[aps]{revtex4}

\usepackage{amsthm}
\usepackage{amsmath}
\usepackage{amstext}
\usepackage{amssymb}

\newcommand{\sgn}{{\rm sign}}
\newcommand{\ret}{\nonumber \\}
\newtheorem{theorem}{Theorem}
\newtheorem{lemma}{Lemma}

\newtheorem{definition}{Definition}
\newcommand{\be}{\begin{equation}}
\newcommand{\ee}{\end{equation}}

\usepackage{graphicx}

\begin{document}

\title{Locality in Quantum Systems}

\author{Matthew~B.~Hastings}

\affiliation{Microsoft Research, Station Q, Elings Hall, University of California,
Santa Barbara, CA 93106, USA.}

\begin{abstract}
These lecture notes focus on the application of ideas of locality, in particular Lieb-Robinson bounds, to quantum many-body systems.
We consider applications including correlation decay, topological order, a higher dimensional Lieb-Schultz-Mattis theorem, and a nonrelativistic
Goldstone theorem.  The emphasis is on trying to show the ideas behind the calculations.  As a result, the proofs are only sketched with
an emphasis on the intuitive ideas behind them,
and in some cases we use techniques that give very slightly weaker bounds for simplicity.

This is a preliminary version of the lecture notes, with the goal of getting the notes out close to the end of the school.  Comments welcome.
\end{abstract}
\maketitle

\section{Introduction and Notation}
The basic problem studied in quantum many-body theory is to find the properties of the ground state of a given Hamiltonian.
Expressed mathematically, the Hamiltonian $H$ is a Hermitian matrix, and the ground state, which we write $\psi_0$, is the eigenvector
of this matrix with the lowest eigenvalue.  The properties we are interested in studying are expectation values of various observables:
given a Hermitian matrix $O$, we would like to compute the expectation value $\langle \psi_0, O \psi_0 \rangle$.  A note on notation: we write an inner product of two vectors as $\langle v,w \rangle$, rather than the notation $\langle v | w \rangle$, as the first notation is more common
in math and also will be more clear given the large number of absolute value signs we will also use later).  Sometimes we will write
the expectation value in the ground state simply as $\langle O \rangle$.

The above paragraph sets out some of the mathematics of quantum mechanics, but it must seem quite dry.  Someone reading that who is
not familiar with quantum theory could be excused for thinking that the problem is essentially one of linear algebra, and that the
basic tool employed by quantum physicists is a linear algebra package to diagonalize large matrices.  In fact, while the method of
``exact diagonalization" on a computer is an important technique in studying quantum systems, it is only a small part of how physical problems
are studied.  The feature that distinguishes the study of quantum many-body systems is {\it locality} of interactions.  Consider
a typical Hamiltonian, such as the one-dimensional transverse field Ising model:
\be
\label{tfi}
H=-J \sum_{i=1}^{N-1} S^z_i S^z_{i+1} + B \sum_{i=1}^{N} S^x_i.
\ee
The very notation we employ to describe this Hamiltonian implicitly assumes local interactions.  To be more precise, throughout these
notes, we have in mind quantum systems on a finite size lattice.  We associate a $D$ dimensional Hilbert space with each lattice
site, and the Hilbert space of the whole system is the tensor product of these spaces.
We use $N$ to represent the number of lattice sites, so that the Hilbert space on which a Hamiltonian such as
(\ref{tfi}) is defined is $D^N$ dimensional.  A term such as $S^z_i S^z_{i+1}$ is a short-hand notation for the term
$I_1 \otimes I_2 \otimes ... \otimes I_{i-1} \otimes S^z_i \otimes S^z_{i+1} \otimes I_{i+2} \otimes ... \otimes I_N$, where
$I_j$ is the identity operator on site $j$.  This local structure of the interaction terms will greatly constrain the properties of $\psi_0$,
in particular if there is a spectral gap.

We will usually use symbols $i,j,k,...$ to denote lattice sites and we use letters $X,Y,Z,...$ to denote sets of lattice sites.
We use $\Lambda$ to denote the set of all lattice sites.  We use $|X|$ to denote the cardinality of a set $X$.

We say that an operator $O$ is ``supported on set $A$" if we can write $O$ as a tensor product of two operators:
\be
O=I_{\Lambda\setminus A} \otimes P,
\ee
where $I_{\Lambda\setminus A}$ is the identity operator on the sites not in set $A$ (note that $\Lambda\setminus A$ denotes the set of sites
not in $A$) and $P$ is some operator defined on the
$D^{|A|}$ dimensional Hilbert space on set $A$.  For example, the operator $S^z_i S^z_{i+1}$ is supported on the set of sites
$\{i,i+1\}$.  Colloquiually, instead of saying that $O$ is supported on set $A$, one often hears that $O$ ``acts on set $A$", although
we will avoid that terminology.

Finally, we define $\Vert O \Vert$ to represent the ``operator norm" of an operator $O$.  If $O$ is Hermitian, the operator norm is
equal to the absolute value of the largest eigenvalue of $O$.  For arbitrary operators $O$, we define
\be
\Vert O \Vert = {\rm max}_{\psi,|\psi|=1} |O\psi|,
\ee
where the maximum is taken over vectors $\psi$ with norm $1$, and $|O\psi|$ denotes the norm of the vector $O\psi$.

Using two simple assumptions, that the Hamiltonian has local interactions and that the Hamiltonian has a spectral gap, we
will be able to prove a wide variety of results about the ground state of the Hamiltonian.  We will consider two different cases of
a spectral gap.  In one case, we consider a Hamiltonian with a unique ground state, $\psi_0$, which has energy $E_0$, with the next
lowest energy state having energy $E_1$ with $E_1-E_0\geq \Delta E$.  Here, $\Delta E$ is called the ``spectral gap".  In the other
case, we consider a Hamiltonian with several degenerate or approximately degenerate low energy states, $\psi_0^a$, for $a=1,...,k$.  Here, $k$
is the number of different low energy states.  Let these states have energies $E_0^a$.  Then, assume that there is a gap $\Delta E$ separating
these ground states from higher energy states.  That is, the $k+1$-st smallest eigenvalue of the Hamiltonian is greater than or equal to
$E_0^a+\Delta E$ for all $a=1,...,k$.

We will begin by considering properties of correlations in these systems, and prove an exponential decay of correlation functions.
We then consider how the ground state of the system changes under a change in the Hamiltonian, and use this to prove a non-relativistic
variant of Goldstone's theorem.  Finally, we apply these techniques to more interesting systems with topological order, beginning with
the Lieb-Schultz-Mattis theorem.
However, before we can do any of this, we need to more precisely define the locality properties of the Hamiltonian and to prove
a set of bounds on the propagation of information through the system called ``Lieb-Robinson bounds".

\section{Locality and Lieb-Robinson Bounds}
To define locality, we need a notion of distance.  We introduce a metric on the lattice, ${\rm dist}(i,j)$.  For example, on a square
lattice one may consider a Manhattan metric or a Euclidean metric.  In one dimension on a system with open boundary conditions, it is natural to consider the metric ${\rm dist}(i,j)=|i-j|$, while for a one dimensional system with periodic boundary conditions and $N$ sites it is natural to consider the metric ${\rm dist}(i,j)={\rm min}_n |i-j+nN|$, where the minimum ranges over integers $n$.
We define the
distance between two sets $A,B$ to be
\be
{\rm dist}(A,B)={\rm min}_{i\in A,j\in B} {\rm dist}(i,j).
\ee
We define the diameter of a set $A$ by
\be
{\rm diam}(A)={\rm max}_{i,j \in A} {\rm dist}(i,j).
\ee

We consider Hamiltonians
\be
H=\sum_Z H_Z,
\ee
where $H_Z$ is supported on a set $Z$.  We will be interested in studying problems in which $\Vert H_Z \Vert$ decays rapidly
with the diameter of the set $Z$.  For example, in the one-dimensional transverse field Ising model Hamiltonian (\ref{tfi}), with
metric ${\rm dist}(i,j)=|i-j|$, the
only terms that appears have diameter $0$ or $1$ (the magnetic field term has diameter zero and the Ising interaction has
diameter one), so $\Vert H_Z \Vert=0$ for ${\rm diam}(Z)>1$.  These interactions are ``finite range"; we will also be able to consider
cases in which the interactions have a slower decay, such as an exponential decay (one can prove various results about power law decay
also, which we will not consider in these notes).

Note that the reason one chooses one particular metric over another is to make the terms in the Hamiltonian local with respect to that
metric.  Note that Hamiltonian (\ref{tfi}) has open boundary conditions, in that site $N$ does not interact with site $1$.
Instead, if we considered a one-dimensional transverse field Ising
\be
\label{tfipbc}
H=-J \sum_{i=1}^{N-1} S^z_i S^z_{i+1} -J S^z_N S^z_1 + B \sum_{i=1}^{N} S^x_i,
\ee
with periodic boundary conditions, the term $S^z_N S^z_1$ would have diameter $N-1$ with respect to the metric ${\rm dist}(i,j)=|i-j|$.
However, if we instead consider the metric
${\rm dist}(i,j)={\rm min}_n |i-j+nN|$ suggested above, then all of the interactions have diameter $0$ or $1$ again.

As an aside, one might wonder why we do not just pick the metric ${\rm dist}(i,j)=0$ for all $i,j$.  In this case, all of the interactions
in the Hamiltonian have diameter zero and are therefore ``local".  However, the kinds of results we will prove below have to do
with exponential decay of various quantities with distance (for example, decay of correlations as a function of spacing between operators),
and all of the results would become trivial in the case of the metric ${\rm dist}(i,j)=0$.  Thus, we want to pick a metric such that
our Hamiltonian is local, but such that the set  of sites $\Lambda$ has a large diameter.

It is not necessary that the system be defined on a regular lattice.  For example, we could imagine a graphical structure to the
interactions.  Suppose each site is associated with the vertex of a graph, and the Hamiltonian is a sum of terms $H_Z$ where each
$Z$ contains only two sites and $H_Z$ is non-vanishing only if there is an edge of the graph connecting the sites in $Z$.  Then,
the shortest path metric on the graph gives us a metric for which $H_Z$ is non-vanishing only if $Z$ has diameter $0$ or $1$.

We now consider the Lieb-Robinson bounds.  These are bounds describing time evolution of operators in a local Hamiltonian.
While most of our interest in these notes is in static properties of systems (such as correlation functions at a given instant in time),
it turns out that an understanding of the dynamical properties as a function of time is very useful to prove results about the statics.
The bounds were first proven in \cite{lr1}.  Then, a proof which does not involve the dimension of the Hilbert space dimension on
a given site was given in \cite{hast-lsm}.  More general lattices were considered in \cite{ns}, but only using a proof which depended
on the Hilbert space dimension again.
The dimension independent proof for arbitrary lattices was given in \cite{koma} and this is what we follow.

Various sorts of Lieb-Robinson bounds can be proven.  We consider first the case of exponentially decaying interactions in some detail,
then we consider either kinds of decay in the interaction (either finite range interactions or other types of decay).
We define the time-dependence of operators by the Heisenberg evolution:
\be
O(t)\equiv \exp(i H t) O \exp(-i H t).
\ee
Then,
\begin{theorem}
Suppose for all sites $i$, the following holds:
\begin{equation}
\label{Hdecay}
\sum_{X\ni i}
\Vert H_X\Vert|X|\exp[{\mu\>{\rm diam}(X)}]\le s<\infty,
\end{equation}
for some positive constants $\mu,s$.
Let $A_X,B_Y$ be operators supported on sets
$X,Y$, respectively.  
Then, if ${\rm dist}(X,Y)>0$,
\begin{eqnarray}
\label{lrbound}
\Vert [A_X(t),B_Y]\Vert&\leq&
 2\Vert A_X\Vert \Vert B_Y\Vert \sum_{i \in X} \exp[-\mu\>{\rm dist}(i,Y)]\left[e^{2s|t|}-1\right]
\\ \nonumber
&\leq &
 2\Vert A_X\Vert \Vert B_Y\Vert |X| \exp[-\mu\>{\rm dist}(X,Y)]\left[e^{2s|t|}-1\right].
\end{eqnarray}
\end{theorem}

Before proving the theorem, let us describe the physical meaning of this theorem.
Suppose $A_X$ is some operator.  Let $B_l(X)$ denote the ball of radius $l$ about
set $X$.  That is, $B_l(X)$ is the set of sites $i$, such that ${\rm dist}(i,X)\leq l$.
Following \cite{bhv}, define
\be
\label{localized}
A_X^l(t)=\int {\rm d}U U A_X(t) U^\dagger,
\ee
where the integral is over unitaries supported on the set of sites $\Lambda \setminus  B_l(X)$ with the Haar measure.
Then, $A_X^l$ is supported on $B_l(X)$.  Since $U A_X(t)  U^\dagger=A_X(t)+U[A_X(t),U^\dagger]$, we have
\be
\Vert A_X^l(t)-A_X(t) \leq \int {\rm d}U \Vert [A_X ,U] \Vert.
\ee
Using the Lieb-Robinson bound (\ref{lrbound}) to bound the right-hand side of the above equation,
we see that $A_X^l(t)$ is exponentially close to $A_X$ if $l$ is sufficiently large compared to $2st/\mu$.  Thus,
to exponential accuracy, we can approximate a time-evolved operator such as $A_X(t)$ by an operator supported on the set
$B_l(X)$.  That is, the ``leakage" of the operator outside the light-cone is small.

Another comment regarding the assumptions on the theorem: because the term $\exp[\mu {\rm diam}(Z)]$ grows exponentially in
$Z$, we need the norm of the terms $H_Z$ to decay exponentially in $Z$.  On finite dimensional lattices (such as a hypercubic
lattice), the cardinality of $Z$, $|Z|$, is bounded by a power in the diameter of $Z$.  So, on such lattices, if the terms $H_Z$
in the Hamiltonian decay exponentially in ${\rm diam}(Z)$, we can find a $\mu$ such that the assumptions of the theorem
are satisfied.  By expressing the theorem as we have, however, the theorem can be applied to models defined on, for example,
arbitrary graphs, where the cardinality of a set might not be bounded by a power of its diameter.

One final remark: we have implicitly assumed that all of the operators are bosonic, in that operators supported on disjoint sets commute
with each other.  One can straightforwardly generalize all of this to  the case of fermionic operators also, so that two fermionic
operators which are supported on disjoint sets anti-commute but that two bosonic operators, or one bosonic and one fermionic operator, commute
on disjoint sets.  In this case, one can instead prove a bound on the anti-commutator of two fermionic operators at different times.

\subsection{Proof of Lieb-Robinson Bound}
The proof of the Lieb-Robinson bound we now give can be straightforwardly adapted to time-dependent Hamiltonians, though we only
present the proof in the time-independent case for simplicity of notation.

Recall that we assume that $|\Lambda|$ is finite.
If it is necessary to consider the infinite volume limit,  
we take the limit after deriving the desired Lieb-Robinson bounds which 
hold uniformly in the size of the lattice. 
Our essential tool is the series expansion (\ref{CBseries}) below 
for the commutator $[A(t),B]$.   Let $A$ be supported on $X$ and $B$ be supported on $Y$.
We assume $t>0$ because negative $t$ can be treated in the same way. 
Let $\epsilon=t/N$ with a large positive integer $N$, and let 
\begin{equation}
t_n=\frac{t}{N}n\quad\mbox{for}\ n=0,1,\ldots, N. 
\end{equation}
Then we have 
\begin{equation}
\left\Vert[A(t),B]\right\Vert-\left\Vert[A(0),B]\right\Vert 
=\sum_{i=0}^{N-1}\epsilon\times
\frac{\left\Vert[A(t_{n+1}),B]\right\Vert-\left\Vert[A(t_n),B]\right\Vert}{\epsilon}.
\label{sumid} 
\end{equation}
In order to obtain the bound (\ref{Commnormbound}) below, 
we want to estimate the summand in the right-hand side. 
To begin with, we note that the identity,  
$\left\Vert U^\ast OU\right\Vert=\Vert O\Vert$, holds for any observable $O$ 
and for any unitary operator $U$. Using this fact, we have  
\begin{eqnarray}
\left\Vert[A(t_{n+1}),B]\right\Vert-\left\Vert[A(t_n),B]\right\Vert
&=&\left\Vert[A(\epsilon),B(-t_n)]\right\Vert-\left\Vert[A,B(-t_n)]\right\Vert\ret
&\le&\left\Vert[A+i\epsilon[H_\Lambda,A],B(-t_n)]\right\Vert
-\left\Vert[A,B(-t_n)]\right\Vert+{\cal O}(\epsilon^2)\ret
&=&\left\Vert[A+i\epsilon[I_X,A],B(-t_n)]\right\Vert
-\left\Vert[A,B(-t_n)]\right\Vert+{\cal O}(\epsilon^2)\ret
\label{difnorm}
\end{eqnarray}
with 
\begin{equation}
I_X=\sum_{Z:Z\cap X\ne \emptyset}H_Z,
\label{defIX}
\end{equation}
where we have used 
\begin{equation}
A(\epsilon)=A+i\epsilon[H_\Lambda,A]+{\cal O}(\epsilon^2)
\end{equation}
and a triangle inequality. Further, by using 
\begin{equation}
A+i\epsilon[I_X,A]=e^{i\epsilon I_X}Ae^{-i\epsilon I_X}+{\cal O}(\epsilon^2), 
\end{equation}
we have 
\begin{eqnarray}
\left\Vert[A+i\epsilon[I_X,A],B(-t_n)]\right\Vert
&\le&\left\Vert[e^{i\epsilon I_X}Ae^{-i\epsilon I_X},B(-t_n)]\right\Vert
+{\cal O}(\epsilon^2)\ret
&=&\left\Vert[A,e^{-i\epsilon I_X}B(-t_n)e^{i\epsilon I_X}]\right\Vert
+{\cal O}(\epsilon^2)\ret
&\le&\left\Vert[A,B(-t_i)-i\epsilon[I_X,B(-t_n)]]\right\Vert
+{\cal O}(\epsilon^2)\ret
&\le&\left\Vert[A,B(-t_n)]\right\Vert+\epsilon\left\Vert[A,[I_X,B(-t_n)]]\right\Vert
+{\cal O}(\epsilon^2). 
\end{eqnarray}
Substituting this into the right-hand side in the last line of (\ref{difnorm}), 
we obtain 
\begin{eqnarray}
\left\Vert[A(t_{n+1}),B]\right\Vert-\left\Vert[A(t_n),B]\right\Vert
&\le&\epsilon\left\Vert[A,[I_X,B(-t_n)]]\right\Vert
+{\cal O}(\epsilon^2)\ret
&\le&2\epsilon\Vert A\Vert\left\Vert[I_X(t_n),B]\right\Vert
+{\cal O}(\epsilon^2).
\end{eqnarray}
Further, substituting this into the right-hand side of (\ref{sumid}) and  
using (\ref{defIX}), we have 
\begin{eqnarray}
\left\Vert[A(t),B]\right\Vert-\left\Vert[A(0),B]\right\Vert 
&\le&2\Vert A\Vert\sum_{n=0}^{N-1}\epsilon\times\left\Vert[I_X(t_n),B]\right\Vert
+{\cal O}(\epsilon)\ret
&\le&2\Vert A\Vert\sum_{Z:Z\cap X\ne \emptyset}\sum_{n=0}^{N-1}\epsilon\times
\left\Vert[H_Z(t_n),B]\right\Vert+{\cal O}(\epsilon). 
\end{eqnarray}
Since $H_Z(t)$ is a continuous function of the time $t$ for a finite volume, 
the sum in the right-hand side converges to the integral in 
the limit $\epsilon\downarrow 0$ ($N\uparrow\infty$) 
for any fixed finite lattice $\Lambda$. In consequence, we obtain
\begin{equation}
\left\Vert[A(t),B]\right\Vert-\left\Vert[A(0),B]\right\Vert 
\le 2\Vert A\Vert\sum_{Z:Z\cap X\ne \emptyset}\int_0^{|t|}ds 
\left\Vert[H_Z(s),B]\right\Vert.
\label{Commnormbound} 
\end{equation}

We define 
\begin{equation}
C_B(X,t):=\sup_{A\in{\cal A}_X}\frac{\Vert[A(t),B]\Vert}{\Vert A\Vert},
\label{CBXt}
\end{equation}
where ${\cal A}_X$ is the set of observables supported on
the set $X$. Then we have
\begin{equation}
C_B(X,t)\le C_B(X,0)+2\sum_{Z:Z\cap X\ne \emptyset}\Vert H_Z\Vert
\int_0^{|t|}ds\> C_B(Z,s)
\label{CBXtbound}
\end{equation}
from the above bound (\ref{Commnormbound}).
Assume ${\rm dist}(X,Y)>0$. 
Then we have $C_B(X,0)=0$ from the definition of $C_B(X,t)$, and note that 
\begin{equation}
C_B(Z,0)\le 2 \Vert B \Vert,
\ee
for $Z\cap Y\ne\emptyset$ and
\be
C_B(Z,0))=0
\ee
otherwise.
Using these facts and the above bound (\ref{CBXtbound}) iteratively, 
we obtain 
\begin{eqnarray}
C_B(X,t)&\le&2\sum_{Z_1:Z_1\cap X\ne\emptyset}\Vert H_{Z_1}\Vert
\int_0^{|t|}ds_1\> C_B(Z_1,s_1)\ret
&\le&2\sum_{Z_1:Z_1\cap X\ne\emptyset}\Vert H_{Z_1}\Vert
\int_0^{|t|}ds_1\> C_B(Z_1,0)\ret
&+&2^2\sum_{Z_1:Z_1\cap X\ne\emptyset}\Vert H_{Z_1}\Vert
\sum_{Z_2:Z_2\cap Z_1\ne\emptyset}\Vert H_{Z_2}\Vert
\int_0^{|t|}ds_1\int_0^{|s_1|}ds_2\> C_B(Z_2,s_2)\ret
&\le&2\Vert B\Vert(2|t|)\sum_{Z_1:Z_1\cap X\ne\emptyset,Z_1\cap Y\ne\emptyset}
\Vert H_{Z_1}\Vert\ret
&+&2\Vert B\Vert\frac{(2|t|)^2}{2!}
\sum_{Z_1:Z_1\cap X\ne\emptyset}\Vert H_{Z_1}\Vert
\sum_{Z_2:Z_2\cap Z_1\ne\emptyset,Z_2\cap Y\ne\emptyset}\Vert H_{Z_2}\Vert\ret
&+&2\Vert B\Vert\frac{(2|t|)^3}{3!}
\sum_{Z_1:Z_1\cap X\ne\emptyset}\Vert H_{Z_1}\Vert
\sum_{Z_2:Z_2\cap Z_1\ne\emptyset}\Vert H_{Z_2}\Vert
\sum_{Z_3:Z_3\cap Z_2\ne\emptyset,Z_3\cap Y\ne\emptyset}\Vert H_{Z_3}\Vert+\cdots
\ret
\label{CBseries}
\end{eqnarray}

We now bound each term in Eq.~(\ref{CBseries}), using the assumption on the decay of terms in the Hamiltonian (\ref{Hdecay}).
The first term is bounded by $2 (2|t|) \sum_{i\in X} \exp(-\mu{\rm dist}(i,Y)$.  The second term is bounded by
\be
2 \frac{(2|t|)^2}{2!} \sum_{i\in X} \sum_{Z_1 \ni i} \sum_{j\in Z_1} \sum_{Z_2 \ni j,Z_2\cap Y \neq \emptyset} 1.
\ee
Recall that  ${\rm dist}(i,Y)\leq {\rm dist}(i,j)+{\rm dist}(j,Y)$.  Thus, $\exp[-\mu{\rm dist}(i,Y)] \exp[\mu{\rm dist}(i,j)]\exp[\mu{\rm dist}(j,Y)] \geq 1$.
Thus, the second term is bounded by
\begin{eqnarray}
&&
2 \frac{(2|t|)^2}{2!} \exp[-{\rm dist}(i,Y)] \sum_{i\in X} \sum_{Z_1 \ni i} 
\sum_{j\in Z_1} 
\exp[\mu{\rm dist}(i,j)]
\sum_{Z_2 \ni j,Z_2\cap Y \neq \emptyset}
\exp[\mu{\rm dist}(j,Y)]
\\ \nonumber
&\leq &
2 \frac{(2|t|)^2}{2!} \exp[-{\rm dist}(i,Y)] \sum_{i\in X} \sum_{Z_1 \ni i} 
\sum_{j\in Z_1} 
\exp(\mu{\rm dist}(i,j))
\sum_{Z_2 \ni j,Z_2\cap Y \neq \emptyset}
\exp[\mu{\rm diam}(Z_2)]
\\ \nonumber
&\leq &
2 \frac{(2|t|)^2}{2!} \exp[-{\rm dist}(i,Y)] \sum_{i\in X} \sum_{Z_1 \ni i} 
\sum_{j\in Z_1} 
\exp[\mu{\rm dist}(i,j)] s 
\\ \nonumber
& \leq &
2 \frac{(2|t|)^2}{2!} \exp[-{\rm dist}(i,Y)] \sum_{i\in X} \sum_{Z_1 \ni i} 
|Z_1| \exp[\mu{\rm diam}(Z_1)] s 
\\ \nonumber
& \leq &
\exp[-\mu{\rm dist}(i,Y)]
2 \frac{(2|t|)^2}{2!} s^2.
\end{eqnarray}

Proceeding in this fashion, we bound the $n$-th term in the series (\ref{CBseries}) by
$2 \exp[-\mu {\rm dist}(i,Y)] (2s|t|)^n/n!$.  Adding these together, we arrive at the bound (\ref{lrbound}).

\subsection{The Lieb-Robinson Velocity and Finite-Range Interactions}
The bound (\ref{lrbound}) is not in quite the most convenient form for later use.  The most convenient form of the theorem is that, supposing
the bound (\ref{Hdecay}) holds, then there is a constant $v_{LR}$ depending only on $s,\mu$ such that for
$t\leq {\rm dist}(X,Y)/v_{LR}$, we have
\be
\label{lrboundconv}
\Vert [A_X(t),B_Y]\Vert \leq
\frac{v_{LR} |t|}{l} g(l) |X| \Vert A_X \Vert \Vert B_Y \Vert,
\ee
where $l={\rm dist}(X,Y)$ and $g(l)$ decays exponentially in $l$.
One may choose, for example, $v_{LR}=4s/\mu$ and Eq.~(\ref{lrboundconv}) will follow from Eq.~(\ref{lrbound}).
Similarly, the operator $A_X(t)$ can be approximated by an operator $A_X^l(t)$ supported on the
set of sites within distance $l=v_{LR} t$ of the set $X$ up to an error  bounded by
$\frac{v_{LR} |t|}{l} g(l) |X| \Vert A_X \Vert$.

The reader should note that in almost all applications of the Lieb-Robinson bound, the error term on the right-hand side
of Eq.~(\ref{lrboundconv}) associated with the ``leakage"
outside the light-cone is negligible.  For example, in the next section, we will prove exponential decay of correlation
functions.  The proof will involve summing up various error terms.  The most important error terms will not arise from the
Lieb-Robinson bound.  In fact, for clarity, the reader may wish to pretend that such error terms are zero on the first reading of
any of the proofs in the rest of the paper, and only later worry about how large these errors are.

Other types of interaction decay can be consider similarly.  Suppose the interaction is finite range.  For example, suppose $H_Z=0$ for
${\rm diam}(Z)>R$ for some interaction range $R$.  Then, we can find constants $\mu,s$ such that the assumption (\ref{Hdecay}) is satisfied.
However, we can do even better in the case of finite range interaction.  Suppose each term in the Hamiltonian only has support on a set of two sites, and suppose we have $R=1$ for simplicity and suppose $\Vert H_Z\Vert \leq J$ for some constant $J$.  We again use the series (\ref{CBseries}).  However, we note that we have now bounded $C_B(X,t)$ by a sum over paths on the lattice starting at sites in $X$ and ending
at sites in $Y$, with each path weighted by $(2 J |t|)^l/l!$, where $l$ is the length of the path.  Such a weighted sum over paths
is a well-studied problem in statistical physics.  In this case, one finds that $g(l)$ decays faster than exponentially in $l$
(roughly, $g(l)$ is $\exp(-{\rm const}*l^2)$ as one may verify).  Other cases of finite range interaction can be handled similarly.

\subsection{Other Types of Decaying Interactions}
\label{otherdecay}
One may also consider other types of decaying interaction, slower than exponential.  This subsection will be useful in considering
quasi-adiabatic continuation later, but may be skipped on first reading.

The important things to remember from the calculations in the rest of the section
are that one can treat decay other than exponential, and that decays which are slower than exponential
decay give rise to Lieb-Robinson bounds with error terms that also decay slower than exponential.  That is, we find
error terms on the right-hand side of the Lieb-Robinson bounds consisting of a function which is exponentially
growing in time, multiplied by a function which decays slower than exponentially in space.  As a result, these types of
decay do {\it not} give rise to a Lieb-Robinson velocity; for example, a function $\exp(t)/l^3$ is only small for $t$ which
is logarithmically large in $l$, while a function $\exp(t) \exp(-\sqrt{l})$ is only small for $t$ of order $\sqrt{l}$.

A further important  thing to remember is that the sum over sites in
Eq.~(\ref{CBseries}) can worsen the bounds beyond what one might expect.  That is, the decay in the error bound may be not as rapid as the decay in the interactions.  For example, if we have a two-dimensional lattice, with an interaction $H$ that is a sum of
$H_Z$ with each $H_Z$ having $|Z|=2$ (i.e., each term in the interaction acts on only two sites), with $\Vert H_Z \Vert \sim 1/{\rm diam}(Z)$,
then we find that
if the distance between a site $i$ and a set $Y$ is equal to $l$, then the first term in Eq.~(\ref{CBseries}) is of order
$1/l$.  The second term is of order $\sum_{j} (1/{\rm dist}(i,j)) (1/{\rm dist}(j,Y))$ where the sum is over sites $j$.  However, one may see 
(by replacing the sum by an integral) that the second term diverges logarithmically as the lattice size tends to infinity.  So, in fact
in this case no Lieb-Robinson bound can be proven.  One needs a faster power-law decay than this to prove the Lieb-Robinson bound
for a two-dimensional lattice.  General consideration of different power-law decays was given in \cite{koma}.

One way of treating these was considered in \cite{mob}.
We define
\begin{definition}
A function $K(l)$ is {\bf reproducing} for a given lattice $\Lambda$ if, for any pair of sites $i,j$ we have
\be
\sum_m K({\rm dist}(i,m)) K({\rm dist}(m,j) \leq \lambda K({\rm dist}(i,j)),
\ee
for some constant $\lambda$.
\end{definition}
For a square lattice in $D$ dimensions and a shortest-path metric, a powerlaw $K(l)\sim l^{-\alpha}$ is reproducing for sufficiently
large $\alpha$.  An exponential decay is {\it not} reproducing.  However an exponential multiplying a sufficiently fast decaying power
is.  Using this definition and Eq.~(\ref{CBseries}), suppose that $\Vert H_Z \Vert \leq K({\rm diam}(Z))$ for some reproducing $K$.
Then, we find that Eq.~(\ref{CBseries}) is bounded by
\begin{eqnarray}
&&2 K({\rm dist}(i,Y)) (2|t|+\frac{(2|t|)^2}{2!} \lambda +
\frac{(2|t|)^3}{3!} \lambda^2 + ... \\ \nonumber
&\leq & 2
K({\rm dist}(i,Y)) \frac{\exp(2\lambda |t|)-1}{\lambda}.
\end{eqnarray}
Note, as mentioned above, that if $K$ decays slower than exponentially, then this function does not lead to a Lieb-Robinson velocity;
the time $t$ at which we have a meaningful bound will grower slower than exponentially in ${\rm dist}(i,Y)$.

A useful trick is that if we start with a function which is not reproducing (such as an exponential) in many cases we can bound
it by a slightly slower decay exponential times a sufficiently fast power law to arrive at a reproducing decay function.

\section{Correlation Decay}
We now consider the decay of correlations in the ground state of a Hamiltonian with a spectral gap and with a Lieb-Robinson bound.  
Such a decay was first proven in \cite{hast-lsm}.  
In addition to the interest in this result for itself, we consider it because it introduces a combination of two techniques
which will be particularly useful.  We combine Lieb-Robinson bounds with a set of tools using the Fourier transform.  Physically,
one may imagine that a spectral gap $\Delta E$ sets a time scale, $\Delta E^{-1}$, and then the Lieb-Robinson bound
allows us to define a length scale $v_{LR} \Delta E^{-1}$.  The techniques we introduce combining the Lieb-Robinson bound
with the Fourier transform make this statement precise.

The statement
of the theorem with a unique ground state\cite{hast-lsm} is
\begin{theorem}
For a quantum lattice system with  a unique ground state and a spectral gap $\Delta E$, and any operators $A_X,B_Y$ supported on
sets $X,Y$, we have
\begin{eqnarray}
\label{red}
\Bigl| \langle \psi_0,A_X B_Y \psi_0 \rangle - 
\langle \psi_0,A_X \psi_0 \rangle
\langle \psi_0,B_Y \psi_0 \rangle \Bigr| & \leq &
C \Bigl\{ \exp(-l\Delta E/2 v_{LR})+{\rm min}(|X|,|Y|) g(l)\Bigr\} \Vert O_A \Vert \Vert O_B \Vert,
\end{eqnarray}
\end{theorem}
for some constant $C$, where $l={\rm dist}(X,Y)$.
We can also consider systems with multiple ground states $\psi_0^a$ as above, with a spectral gap $\Delta E$ separating those states from
the rest of the spectrum.  In this case\cite{locality}, we can instead obtain bound the quantity
$\langle \psi_0^a,A_X B_Y \psi_0^a\rangle-\langle \psi_a^a, A_X P_0 B_Y \psi_0^a$, where
\be
P_0=\sum_a |\psi_0^a\rangle\langle\psi_0^a |
\ee
is the projector onto the ground state subspace.  The bound on this quantity 
$\langle \psi_0^a,A_X B_Y \psi_0^a\rangle-\langle \psi_a^a, A_X P_0 B_Y \psi_0^a$
is equal to the right-hand side of Eq.~(\ref{red}) plus an additional term which is proportional to the energy difference between the ground states, and in fact this result reduces to Eq.~(\ref{red}) in the case of a single ground
state.

Before sketching the proof of the theorem, we discuss the application of the theorem in various settings.
Consider the transverse field Ising model Hamiltonian in the paramagnetic phase, $B>>J$, with spin-$1/2$ on each site.  In this case, there is a unique ground state and a spectral gap, so that all correlations decay exponentially in space.
On the other hand, we may consider the Hamiltonian in the ferromagnetic phase, $J>>B$.  For $B=0$, the model has two exactly degenerate
ground states, corresponding to all spins pointing up or all spins pointing down.  In fact, the Hamiltonian has a particular symmetry.
It commutes with the unitary operator $\prod_i (2 S^x_i)$.  This operator flips all the spins.  So, the eigenstates of the Hamiltonian
can be chosen to be eigenstates of this particular unitary.  To do this, at $B=0$ we may choose the ground states to be symmetric and
anti-symmetric combinations of the states with all spins up or down: 
\be
\psi_{\pm}=\frac{1}{\sqrt{2}} \Bigl(|\uparrow\uparrow\uparrow...\rangle\pm |\downarrow\downarrow\downarrow...\rangle\Bigr).
\ee
  For $B>0$ but $B$ still much smaller than $J$, there starts to be a small splitting (exponentially small in system size) between the
two lowest energy states, and the gap to the rest of the spectrum remains open.
Now, consider a local operator such as $S^z_i$.  We will see what the action of this operator is in the ground state sector.  That is,
we will construct a $2$-by-$2$ matrix $M$ whose matrix elements are the matrix elements of $S^z_i$ in the ground state.
For $B=0$, we have
\be
\langle \psi_+, S^z_i \psi_+\rangle=
\langle \psi_-, S^z_i \psi_-\rangle=0,
\ee
and
\be
\langle \psi_+, S^z_i \psi_-\rangle=
\langle \psi_-, S^z_i \psi_+\rangle=1/2,
\ee
so
\be
M=\begin{pmatrix} 0 & 1/2 \\ 1/2 & 0 \end{pmatrix}.
\ee  
For $B>0$, we instead find that
\be
\label{ordparam}
M=\begin{pmatrix} 0 & m \\ m & 0 \end{pmatrix},
\ee  
where $m$ is the ``order parameter" ($m$ is greater than zero in the ferromagnetic phase and $m<1/2$ for $B>0$).

The nontrivial matrix elements of $S^z_i$ in the ground state sector are what give rise to the long range correlations: the theorem
above shows that $\langle \psi_+, S^z_i S^z_j \psi_+ \rangle$ approaches $m^2$ when ${\rm dist}(i,j)$ gets large.
This behavior will contrast strongly with the topologically ordered case below.

We only sketch the proof in the case of a unique ground state (the reader is invited to consider the case of multiple ground states).
Define $B_Y^+$ to be the positive energy part of $B_Y$.  That is, let $\{\psi_i\}$ be a basis of eigenstates of $H$, with $\psi_i$ having energy $E_i$.
In this basis, we define $B_Y^+$ to have matrix elements
\be
(B_Y^+)_{ij}=(B_Y)_{ij} \theta(E_i-E_j),
\ee
where $\theta(x)$ is the step function: $\theta(x)=1$ for $x>0$, $\theta(0)=1/2$, and $\theta(x)=0$ for $x<0$.
Without loss of generality, assume that $\langle \psi_0,A_X \psi_0\rangle=\langle \psi_0, B_Y \psi_0\rangle=0$.  Then, we find that
\be
\langle \psi_0,A_X B_Y \psi_0\rangle=
\langle \psi_0,A_X B_Y^+ \psi_0\rangle=
\langle \psi_0,[A_X, B_Y^+] \psi_0\rangle,
\ee
since $B_Y^+\psi_0=B_Y\psi_0$ as there are no states with energy less than $\psi_0$.
Our strategy is to construct an approximation to $B_Y^+$, which we call $\tilde B_Y^+$, which has two properties.
First, $\tilde B_Y^+$ has small commutator with $A_X$.  Second,
\be
\label{ap1}
B_Y^+|\psi_0\rangle \approx \tilde B_Y^+ |\psi_0\rangle,
\ee
and
\be
\label{ap2}
\langle \psi_0| \tilde B_Y^+ \approx \langle \psi_0 | B_Y^+=0,
\ee
where the error in the approximation is discussed below.
Then, by making the error in the approximations small, and by making the commutator $[A_X,\tilde B_Y^+]$ small in operator norm, we will be able to
bound
$\langle \psi_0,[A_X, B_Y^+] \psi_0\rangle$ as follows: we first  show (using (\ref{ap1},\ref{ap2})) that $\langle \psi_0, [A_X,B_Y^+] \psi_0 \rangle$ is close to
$\langle \psi_0,[A_X, \tilde B_Y^+] \psi_0\rangle$.  Then, we show that that quantity is small using  a bound on the operator norm of
the commutator.

To do this, we define $\tilde B_Y^+$ by
\be
\label{tildedef}
\tilde B_Y^+ = \frac{1}{2\pi} \lim_{\epsilon\rightarrow 0^+} \int {\rm d}t B_Y(t) \frac{1}{it+\epsilon} \exp[-(t\Delta E)^2/(2q)],
\ee
where $q$ is a constant that we choose later.  Note that as $q \rightarrow \infty$, we find that the Gaussian on the right-hand
side  of Eq.~(\ref{tildedef}) gets broader, and $\tilde B_Y^+$ converges to $B_Y^+$.  So, taking $q$ large will make it easier to
satisfy (\ref{ap1},\ref{ap2}).  Conversely, for $q$ large, we will be able to show that $\Vert [A_X,\tilde B_Y^+] \Vert$ is small, since
for large times $t$ the integral over $t$ in (\ref{tildedef}) is cut off by the Gaussian, and for short times $t$ the commutator
of $A_X$ with $B_Y(t)$ is bounded by the Lieb-Robinson bound.

Using the energy gap, one may show that
\be
\label{two}
\Bigl| B_Y^+|\psi_0\rangle - \tilde B_Y^+ |\psi_0\rangle \Bigr| \leq C \exp(-q/2) \Vert B_Y \Vert.
\ee
To show this, we bound the absolute value of the matrix element $\langle \psi_i,(B_Y^+-\tilde B_Y^+) \psi_0\rangle$ for some $i>0$.
This is equal to
\be
|\langle \psi_i,(B_Y^+-\tilde B_Y^+) \psi_0\rangle|= |(B_Y)_{i0}| \, \Bigl|1-\int {\rm d}t \exp[i (E_i-E_0) t] \frac{1}{it+\epsilon} \exp[-(t\Delta E)^2/(2q)]\Bigr|.
\ee
The integral in the above equation is equal to the Fourier 
transform of the function
$\frac{1}{it+\epsilon} \exp[-(t\Delta E)^2/(2q)]$.  This Fourier transform is the convolution of the Fourier transform of the step
function with a Gaussian.  In the figure, we sketch the Fourier transform of this function, as well as the step function.  For a narrow width of the Gaussian in frequency (which occurs if $q$ is large), the Fourier transform
of the two functions are very close for energy above $\Delta E$.  One can bound the difference in the Fourier transform by $C\exp(-q/2)$.
So, $|(B_Y^+-\tilde B_Y+)\psi_0|^2\leq \sum_{i>0} |(B_Y)_{i0}|^2 \exp(-q)$, so Eq.~(\ref{two}) follows.
Similarly, one can bound the quantity in (\ref{ap2}) in the same way.

\begin{figure}[tb]
\centerline{
\includegraphics[scale=0.7]{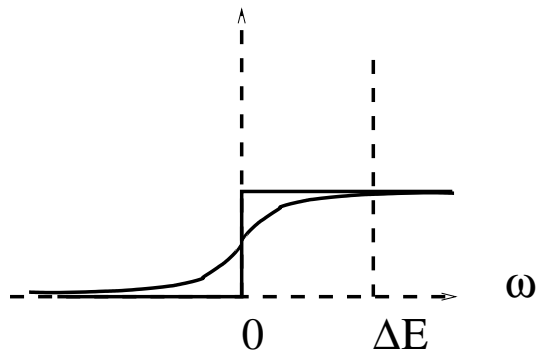}}
\caption{Plot of a step function (solid line) and a sketch of a step function convolved with a Gaussian (solid line).  Dashed line at $\Delta E$ shows that the difference between the functions is small at sufficiently large frequency.}
\end{figure}

One may also bound $\Vert [A_X,\tilde B_Y^+] \Vert$ by a triangle inequality as
\be
\Vert [A_X,\tilde B_Y^+] \Vert \leq
\frac{1}{2\pi} \lim_{\epsilon\rightarrow 0^+} \int {\rm d}t \Vert[A_X,  B_Y(t)] \Vert \, \Bigl| \frac{1}{it+\epsilon} \Bigr| \exp[-(t\Delta E)^2/(2q)].
\ee
To bound this integral, we split the integral over $t$ into times $t$ less than $l/v_{LR}$ and $t>l/v_{LR}$.  For $t<l/v_{LR}$, we use
the Lieb-Robinson bound to bound the integral by a constant times $g(l) |X| \Vert A_X \Vert \Vert B_Y \Vert$.  For $t>l/v_{LR}$, we use
the Gaussian to bound it by a constant times 
\be
\label{one}
\Vert A_X \Vert \Vert B_Y \Vert \exp[-(l\Delta E/v_{LR})^2/2q].
\ee
We now pick $q$ to minimize the sum of terms in (\ref{two},\ref{one}).  Picking $q=l\Delta E/v_{LR}$ is the best possible choice,
and gives the error bound in the theorem.

The most important thing to remember from this sketch is: we combine Lieb-Robinson bounds and Fourier transforms.  We have some function
which is not smooth (like the step function).  We approximate it by some smooth function (the step function convolved with a Gaussian) to
obtain a function whose Fourier transform decays rapidly in time (in this case as $1/t$ times a Gaussian).  This then allows us to
apply the Lieb-Robinson bounds.  The same ideas are behind the technique of quasi-adiabatic continuation later.

\section{Topological Order}
There are many different properties that characterize ``topological order".  Consider a Hamiltonian such as Kitaev's toric code model\cite{toric}.  We will not review this model here, since it is well-explained elsewhere.  However, this model has several unique properties.  On a torus,
the model has 4 exactly degenerate ground states, with an energy gap to the rest of the spectrum.  Surprisingly, however, on other topologies,
the ground state degeneracy is different.  For example, the model has a unique ground state on a sphere.  This contrasts strongly with the
case of the transverse field Ising model mentioned above, which has either 1 ground state (in the paramagnetic phase) or 2 ground states
(in the ferromagnetic phase) independent of the topology of the lattice.  This property of the ground state degeneracy depending upon
topology is one characteristic of a Hamiltonian with topological order.  Another characteristic is certain corrections to the
entanglement entropy\cite{levinwen,kitaevpreskill}.  We will not consider these corrections here.

In fact, we would like to regard both of these properties (the ground state degeneracy and the entropy) as being secondary to one
particular defining feature of topological order which we now explain.  Suppose a model has $k$ different ground states, $\psi_0^a$ for $a=1,...,k$.
Then, given any operator $O$, we can project this operator $O$ into the ground state sector, defining a $k$-by-$k$ matrix $M$
whose matrix elements are
\be
M_{ab}=\langle \psi_0^a, O \psi_0^b \rangle.
\ee
In the case of the toric code on a torus, the matrix $M$ is a 4-by-4 matrix.  Surprisingly, if the operator $O$ is local in the sense
that the diameter of the support of $O$ is sufficiently small compared to $L$, then the matrix $M$ is equal to a constant times the identity matrix.
In the case of the toric code, in fact the matrix $M_{ab}$ is equal to a multiple of the identity matrix so long as the diameter of the support of
$O$ is less than $L/2$, so one can consider operators $O$ which act on a very large number of sites, and yet still the operator
acts just as the identity operator when projected into the ground state sector.
This behavior contrasts strongly with the case of the transverse field Ising model in the ferromagnetic phase, for which the operator
$S^z_i$ acts nontrivially on the ground state sector as discussed above.

If we slightly perturb the toric code, the model remains in the same phase\cite{stability1,stability2}.  The 4 ground states will no longer
be exactly degenerate, but the difference in energy between them will be exponentially small as a function of system size (see \cite{stability1,stability2} for general upper bounds on the energy splitting).  Further, this topological order property will be slightly weakened:
we will instead have a property that if operator $O$ is supported on a set of diameter sufficiently small compared to $L$ (for example, a diameter of at most $L/2$ will suffice for the toric code), then the corresponding matrix $M$ is exponentially close to a multiple the identity matrix
(again, see \cite{stability1,stability2} for general upper bounds).

This property, that local operators are close to a multiple of the identity matrix when projected into the ground state sector
was identified in the Hall effect\cite{wen} and in other topologically ordered states\cite{freedman}.  Note that this property in fact
ensures that the splitting between the ground states is small: the Hamiltonian itself is a sum of local operators, so by this property
of topological order, the Hamiltonian has almost the same expectation value in each of the different ground states.

We can quantify this topological order as in \cite{bhv}.  We say that a system has $(l,\epsilon)$ topological order
if, for any operator $O$ supported on a set of diameter as most $l$, the corresponding matrix $M$ is within $\epsilon$ of a multiple of
the identity.  That is, for some complex number $z$, we have $\Vert M - z I \Vert \leq \epsilon$.

One interesting property of this viewpoint about topological order is that it is a property of a set of states, rather than a property
of a Hamiltonian.  Given any set of orthogonal states, $\psi_0^a$, we can ask whether these states are topologically ordered, independent of whether or
not the states happen to be the ground states of some local Hamiltonian.  

Note that it is {\it not} the case that every state $\psi_0^1$
has some partner state $\psi_0^2$ which gives us a pair of topologically order states.  For example, consider a system of spin-$1/2$
spins and consider the state with
all spins up,
$|\psi_0^1\rangle=|\uparrow\uparrow\uparrow...\rangle$.  There is no state $|\psi_0^2\rangle$ that is orthogonal to $|\psi_0^1\rangle$
such that the two states $\psi_0^a$ for $a=1,2$ are topologically ordered.

\subsection{Topological Order Under Time Evolution}
\label{toute}
Note that we can use the Lieb-Robinson bound to describe the behavior of topological order under time evolution\cite{bhv}.
Suppose a state $\phi$ has 
$(l,\epsilon)$ topological order for some $l,\epsilon$.
Thus, there exists another state $\phi'$ which is the partner of state $\phi$.  
Let us evolve $\phi$ for time $t$ under some Hamiltonian $H$ to obtan $\psi=\exp(-i H t)\phi$ (one can consider also time-dependent
Hamiltonians).
Define
$\psi'=\exp(-i H t)\phi'$.  We now wish to show thhat $\psi,\psi'$ retain some memory of the topological order in $\phi,\phi'$: the
length scale $l$ will be smaller and the error $\epsilon$ will be larger in a way that we can bound quantitatively.

Let $O$ be any local operator supported on a set of diameter $l-m$, for some $m\leq v_{LR} t$.
We project this operator $O$ into the two dimensional space of states spanned by $\psi'$ and $\psi$.  The result is equal to
the projection of the operator
$\exp(-i H t) O \exp(i H t)$ into the space of states spanned by $\phi'$ and $\phi$.  However, the operator
$\exp(-i H t) O \exp(i H t)$ can be approximated, by the Lieb-Robinson bound, by an operator supported on a set of
diameter ${\rm diam}(O)+m\leq l$ up to an error 
$g(m) |X| \Vert A_X \Vert$ as given by Eq.~(\ref{lrboundconv}).
We bound $|X|$ by some constant times $l^d$ for a $d$-dimensional lattice.
Thus, we find that $\psi,\psi'$ have 
$(l-m,\epsilon+
g(m) {\rm const.} l^d) $ topological order for any $m\leq v_{LR} t$.
Thus, topological order cannot be completely destroyed in a short-time.  For example, suppose that the interactions in the
Hamiltonian are such that $g(l)$ decays exponentially in $l$, and suppose
that the initial states $\phi,\phi'$ have $(l,\epsilon)$ topological
order with $l=L/2$ and $\epsilon$ exponentially small in $L$.  Then, choosing $m=L/4$, we find that for times up to $L/(4v_{LR})$, the
states $\phi,\phi'$ have $(L/4,\epsilon')$ topological order, with $\epsilon'$ still being exponentially small in $L$.

Conversely, topological order also cannot be produced in a short
time.  If we start with a state such as
$|\psi_0^1\rangle=|\uparrow\uparrow\uparrow...\rangle$ and  evolve for time $t$ under some Hamiltonian $H$, we cannot produce $(l,\epsilon)$ topological order for $l$ large compared to $v_{LR} t$ and $\epsilon$ small.

\section{Fourier Transforms}
As seen by our analysis of correlation functions, Fourier transforms play a key role in the application of techniques of Lieb-Robinson
bounds to quantum many-body systems.  In this section, we collect a few useful facts about Fourier transforms.

First, the Fourier transform of a Gaussian is a Gaussian.  There is an ``uncertainty principle" at work here: the narrower the
Gaussian in time, the wider the Gaussian in frequency, and vice-versa.

Second, the Fourier transform of the product of two functions is the convolutions of their Fourier transform.  This was used,for example, to estimate
the term $\exp(-q/2)$ in the calculation of correlation function decay.

While we have noticed the general principle that there is a tradeoff between the spread of a function in time and in frequency, it
will be useful in some applications to have functions which have compact support in frequency.  That is, we would like to find
a function $\tilde g(\omega)$ which vanishes for $|\omega|>1$, with $\tilde g(0)=1$, and such that the Fourier transform
$g(t)$ decays as rapidly in time as possibly.
In the classic paper\cite{note}, it is shown how to construct such functions $g(t)$ such
that
\be
|g(t)|\leq {\cal O}(\exp(-|t| \epsilon(|t|))),
\ee
for {\it any} monotonically decreasing positive function $\epsilon(y)$ such that
\be
\int_1^{\infty} \frac{\epsilon(y)}{y} {\rm d}y
\ee
is convergent.  Further, it was shown that this is the optimal possible decay.
For example, the function $\epsilon(y)$ may be chosen to be
\be
\epsilon(y)=1/\log(2+y)^2.
\ee
Thus, this function $g(t)$ has
so-called ``subexponential decay"\cite{subexp}.  A function $f(t)$
is defined to have subexponential decay if,
for any $\alpha<1$,
$|f(t)| \leq C_{\alpha} \exp(-t^\alpha)$, for some $C_{\alpha}$ which depends on $\alpha$.

Thus, while we cannot quite obtain exponential decay in time and compact support in frequency, we can come very close to it,
obtaining functions which ``almost decay exponentially".

We will find it useful also in the next section to have a function $\tilde F(\omega)$ such that $\tilde F(\omega)=1/\omega$ for
$|\omega|\geq 1$ and such that $\tilde F(\omega)$ is odd and the Fourier transform $F(t)$ decays rapidly in time.  We can do this
using the functions $g(t)$ above as follows.
First, we assume without loss of generality that the function $g(t)$ above is an even function of $t$ (if not, simply take
the even part of the function).  Then,
define the even function $f(t)$ by
\be
f(t)=\delta(t)-g(t),
\ee
where $\delta(t)$ is the Dirac $\delta$-function (note that $f(t)$ is thus a distribution rather than a function).
Thus that the Fourier transform $\tilde f(\omega)$ has the property that
$\tilde f(\omega)=0$ for $\omega=0$ and $\tilde f(\omega)=1$ for $|\omega|\geq 1$.
Then, define
$F(t)$ by
\be
\label{da}
F(t)=\frac{i}{2}\int {\rm d}u f(u) \sgn(t-u),
\ee
where $\sgn(t-u)$ is the sign function: $\sgn(t-u)=1$ for $t>u$, $\sgn(t-u)=-1$ for $t<u$, and $\sgn(0)=0$
 (since we convolve $f(u)$ against $\sgn(t-u)$, the resulting $F(t)$ is
a function, rather than a distribution).
We now show the time decay of $F(t)$ and we show that
the Fourier transform $\tilde F(\omega)$ is equal to $-1/\omega$ for $|\omega|\geq 1$, as desired (this calculation is directly from
\cite{mob}).

\begin{lemma}
Let $F(t)$ be as defined in \ref{da}.  Let $\tilde F(\omega)$ be the Fourier transform of $F(t)$.  Then,
\be
|F(t)|\leq |\int_{|t|}^{\infty} f(u)  {\rm d}u|,
\ee
and
\be
\tilde F(\omega)=
\frac{-1}{\omega} \tilde f(\omega).
\ee
\begin{proof}
Assume, without loss of generality, that $t\geq 0$.  Then, we have
$|F(t)|\leq |\int_{t}^{\infty} f(u)  {\rm d}u|/2+
|\int_{-\infty}^{t} f(u) {\rm d}u|/2$.  Since $\tilde f(0)=0$, we have
$|\int_{-\infty}^{t} f(u) {\rm d}u|=
|\int_{t}^{\infty} f(u)  {\rm d}u|$.  Thus,
$|F(t)|\leq |\int_{t}^{\infty} f(u)  {\rm d}u|$.

We have
\be
\tilde F(\omega)=
\frac{i}{2}
\int {\rm d}t \exp(i \omega t)
\int {\rm d}u f(u) \sgn(t-u).
\ee
Integrating by parts in $t$, we have
\begin{eqnarray}
\tilde F(\omega)&= &
\frac{-1}{\omega} \int {\rm d}t \exp(i \omega t)
\int {\rm d}u f(u) \delta(t-u) \\ \nonumber
&=&
\frac{-1}{\omega} \tilde f(\omega).
\end{eqnarray}
Note that $\lim_{t\rightarrow \pm \infty}
\Bigl( \int {\rm d}u f(u) \sgn(t-u)\Bigr)=0$, so the contributions to the integration by parts from the upper and lower limits
of integration vanish.
\end{proof}
\end{lemma}

Finally, given the decay of $f(t)$, it follows that $F(t)$ decays subexponentially also.

A side note: we can also use this idea of compact supported functions in the correlation decay calcuation done previously.
Using the functions $\tilde g(\omega)$ described above, we can construct, for example, a family of functions $\tilde f(\omega,\epsilon)$ such that $\lim_{\epsilon\rightarrow 0}
\tilde f(\omega,\epsilon)=1$ for $\omega \geq 1$ and $\tilde f(\omega,\epsilon)=0$ for $\omega\leq -1$, and with the Fourier transform $g(t,\epsilon)$
decaying subexponentially at large times.  We construct this family by taking the Fourier transform of the function $g(t)/(it+\epsilon)$; the
Fourier transform of this function converges, as $\epsilon\rightarrow 0$, to the convolution of the Fourier transform of $f(t)$ with a step function.
This family of functions still is singular in the limit $t,\epsilon \rightarrow 0$
(a singularity like $1/(it+\epsilon)$, just as we encountered in the Gaussian function $\exp[-(t\Delta E)^2/2q]/(it+\epsilon)$ in the
correlation decay calculation).
Such a singularity is in fact unavoidable given the large $\omega$ behavior of the function.
This approach will
not give quite as tight bounds on the correlation decay (the bounds will be subexponential rather than exponential), but the calculation is a little
simpler since we will have only error terms involving the bound on the commutator $\Vert [A_X,\tilde B_Y^+] \Vert$, while
with these compactly supported functions the difference $\tilde B_Y^+\psi_0-B_Y^+ \psi_0$ will vanish.

\section{Quasi-Adiabatic Continuation}
We now consider the problem of how the ground state of a local Hamiltonian changes as a parameter in the Hamiltonian is
changed.  Suppose we have a parameter dependent Hamiltonian, $H_s$, where $s$ is some real number.  Suppose that $H_s=\sum_Z H_Z(s)$, with
$H_Z(s)$ being differentiable.  Suppose further that we have uniform bounds on the locality properties of the Hamiltonian (for example, for
all $s$, the exponential decay (\ref{Hdecay}) holds or some other similar assumption holds uniformly in $s$).
The main idea of this section can be summarized in a single sentence as follows: if such a Hamiltonian $H_s$ has a lower bound
on the spectral gap which is uniform in $s$ and has a unique ground state, $\psi_0(s)$, then we can define a Hermitian operator, called the quasi-adiabatic continuation operator ${\cal D}_s$, which
is local (in some slightly weaker sense), such that 
either $\partial_s \psi_0(s)= i {\cal D}_s \psi_0(s)$ or
$\partial_s \psi_0(s)\approx i {\cal D}_s \psi_0(s)$ (whether we want exact or approximate equality depends upon the application).
We will also present a generalization to the case of multiple ground states, and use this result to prove a Goldstone theorem.
Finally, we will use these ideas to discuss what we mean by a ``phase" of a quantum system, and to explore the stability of topological
order under perturbations.

In the next section, we will the quasi-adiabatic continuation operator defined in this section to evolve states along paths in parameter
space such that {\it the Hamiltonian does not necessarily have a spectral gap for} $s>0$.  This continuation along these paths will allow us to
prove a higher dimensional Lieb-Schultz-Mattis theorem.  The fact that we continue along paths which might not have a gap is in fact
essential to the proof of that theorem: the results in this section show that continuing along a gapped path implies that one
remains in the ground state, while our goal in the next section is to construct a state which is {\it different } from the ground state but
still low energy to prove a  variational result.

We now define the quasi-adiabatic continuation operator:
\begin{definition}
Given a parameter-dependent Hamiltonian, $H_s$, an operator $O$, and function $F(t)$,
we define
the {\bf quasi-adiabatic continuation operator} to  be the operator
${\cal D}(H_s,O)$ defined by
\begin{eqnarray}
\label{cordef}
i{\cal D}(H_0,O)&=&\int
F(\Delta E t) \exp(i H_s t) O \exp(-i H_s t)
{\rm d}t,
\end{eqnarray}
where $F$ is an odd function of time so that ${\cal D}$ is Hermitian.
\end{definition}
Note that since $F$ is odd, its Fourier transform $\tilde F(\omega)$ obeys
\be
\label{Berryprop}
\tilde F(0)=0,
\ee
which will be useful in discussions of Berry phase later.

Given a parameter dependent Hamiltonian $H_s=\sum_Z H_Z(s)$,
we define
\be
{\cal D}_s={\cal D}(H_s,\partial_s H_s).
\ee
We also sometimes write ${\cal D}^Z_s={\cal D}(H_s,\partial_s H_Z(s))$, so that
\be
\label{de}
{\cal D}_s=\sum_Z {\cal D}^Z_s.
\ee

We will use two different types of functions $F(t)$ in the definition of the quasi-adiabatic continuation operator.
The first type of function is the function $F(t)$ constructed in the previous section, such that the Fourier transform of
$F(t)$ obeys $\tilde F(\omega)=-1/\omega$ for $|\omega|\geq 1$.  This will give, as we now show,
\be
\partial_s \psi_0(s)= i {\cal D}_s \psi_0(s).
\ee
We call the quasi-adiabatic continuation operator arising from such a function $F(t)$ an ``exact quasi-adiabatic continuation
operator".  The second type of function $F(t)$ will lead to only approximate equality
\be
\partial_s \psi_0(s)\approx i {\cal D}_s \psi_0(s),
\ee
and we call this the ``Gaussian quasi-adiabatic continuation operator".

Let us first show
\be
\partial_s \psi_0(s)= i {\cal D}_s \psi_0(s)
\ee
in the case of an exact quasi-adiabatic continuation operator.
We have
\begin{eqnarray}
\label{exact}
i {\cal D}_s \psi_0(s)&=& 
\int F(\Delta E t) \exp(i H_s t) \Bigl( \partial_s H_s \Bigr) \exp(-i H_s t)
{\rm d}t \psi_0(s) \\ \nonumber
&=& \sum_{i\neq 0} |\psi_i(s)\rangle\langle\psi_i(s)|
\int F(\Delta E t) \exp(i H_s t) \Bigl( \partial_s H_s \Bigr) \exp(-i H_s t)
{\rm d}t \psi_0(s) \\ \nonumber
&=& \sum_{i\neq 0} |\psi_i(s)\rangle
\langle\psi_i(s), \Bigl( \partial_s H_s\Bigr) \psi_0(s) \rangle
\int F(\Delta E t) \exp[i (E_i(s)-E_0(s)) t] {\rm d}t \\ \nonumber
&=& \sum_{i\neq 0} \frac{1}{E_0(s)-E_i(s)} |\psi_i(s)\rangle
\langle\psi_i(s), \Bigl( \partial_s H_s\Bigr) \psi_0(s) \rangle \\ \nonumber
&=& \partial_s \psi_0(s),
\end{eqnarray}
where $\psi_i(s)$ for  $i>0$ denote excited states of the Hamiltonian with energy $E_i(s)$.  The  second line follows
by inserting the identity as $\sum_i |\psi_i(s)\rangle\langle\psi_i(s)|$, and noting that property \ref{Berryprop} implies that the term with
$i=0$ is absent.  The third line follows by using the fact that the $\psi_s(s)$ are eigenstates.  The fourth line follows from
the properties of $F(t)$.  The fifth line is ordinary perturbation theory.

In the case of multiple ground states, one can generalize this result as follows.  We instead find that if we have ground states
$\psi_0^a(s)$ which are not exactly degenerate, then
\be
\partial_s \psi_0^a(s)= i {\cal D}_s \psi_0(s) + \sum_b Q_{ab} \psi_0^b(s),
\ee
where $Q_{ab}$ are the matrix elements of some anti-Hermitian matrix $Q$ (if the ground states are exactly degenerate, then
$\partial_s \psi_0^a(s)$ may be ill-defined).
There is an important Berry phase property: if the ground states are exactly degenerate, then the Berry phase arising from the quasi-adiabatic evolution is the
same as the usual non-Abelian Berry phase\cite{goldstone}, while small corrections to this result occur if there is ground state splitting.
Similarly, if $P_0(s)$ is the projector onto the ground state sector of $H_s$, then
\be
\label{P0change}
\partial_s P_0(s)=i [{\cal D}_s,P_0(s)].
\ee

The other important property that the quasi-adiabatic continuation  operator, ${\cal D}_s$ has, in addition to (\ref{exact}), is that
it is local.  Let us assume that $\Vert \partial_s H_Z(s)\Vert$ is bounded by a constant times $\Vert H_Z(s)\Vert$ to fix a normalization
on how rapidly the Hamiltonian changes.
Then if the original Hamiltonian $H$ has a superpolynomial decay in its interactions (so that $\Vert H_Z \Vert$
decays superpolynomially in ${\rm diam}(Z)$), and the lattice is finite dimensional, then ${\cal D}_s$ also has a superpolynomial decay:
we can write ${\cal D}_s=\sum_Z D_Z(s)$, where $D_Z(s)$ is supported on $Z$, with $\Vert D_Z \Vert$ decaying superpolynomially
in ${\rm diam}(Z)$.  Note that $D_Z(s)$ is {\it not} the same thing as ${\cal D}^Z_s$.  Indeed, ${\cal D}^Z_s$ is not supported on $Z$.

To prove the locality of $D_Z$, one uses the Lieb-Robinson bounds and the superpolynomial decay of the function $F(t)$.  Before
 sketching the proof, let us give the basic idea.  
  Consider a given
${\cal D}^Z_s$.  
in integral in Eq.~(\ref{cordef}) is small, while at short time we can approximate $\exp(i H_s t) (\partial_s H_Z(s)) \exp(-i H_s t)$
by an operator supported near $Z$.
More precisely, we will decompose ${\cal D}^Z_s=\sum_{l=0}^{\infty} O_l(Z)$, where $O_l(Z)$ is supported on the set of sites within distance $l$ of
$Z$ as follows.  We define
\be
O_0(Z)=\int 
\int
F(\Delta E t) \partial_s H_Z(s))^0(t)
{\rm d}t,
\ee
where, following Eq.~(\ref{localized}),  
$(\partial_s H_Z(s))^0(t)$ denotes an approximation to 
$(\partial_s H_Z(s))(t) \equiv  \exp(i H_s t) (\partial_s H_Z(s)) \exp(-i H_s t)$ which is localized on set $Z$.
We define, for $l>0$,
\be
O_l(Z)=\int 
F(\Delta E t) 
\Bigl((\partial_s H_Z(s))^l(t)-
(\partial_s H_Z(s))^{l-1}(t)\Bigr)
{\rm d}t.
\ee
Summing over $l$ recovers the desired result.
Now, we define $D_Z(s)$ to be the sum over $Y$ of the $O_l(Y)$ which are supported on $Y$.

As mentioned, one can also consider Gaussian quasi-adiabatic continuation operators.  These were the first type of quasi-adiabatic continuation
operators considered\cite{hast-lsm}, while the exact operators will considered later\cite{osborne}.  The Gaussian quasi-adiabatic
continuation operators in some cases lead to tighter bounds.  For example, in the Lieb-Schultz-Mattis theorem, they lead to tighter
bounds than the exact operators due to the faster time decay.  On the other hand, in many cases the exact operators are much more
convenient.  {\it At this point, we make a deliberate choice due to the nature of these lecture notes.  Rather than consider the
Gaussian operators in detail (which leads to an enormous number of triangle inequalities in the actual calculations, potentially 
obscuring the physics), we will only use the exact quasi-adiabatic continuation operators.}  This will lead to slightly less tight results
in many cases, but the improvement in clarity (and generality in considering topological phases later), seems well worth it.

\subsection{Lieb-Robinson Bounds for Quasi-Adiabatic Continuation}
One particular advantage of considering the exact quasi-adiabatic continuation operators is that we have a Lieb-Robinson bound
for them.  We have shown that the norm of the terms $D_Z(s)$ decays superpolynomially in ${\rm diam}(Z)$ (indeed, it decays
subexponentially if the original Hamiltonian is a finite dimensional lattice with exponentially decaying interactions).  This implies
(see the previous discussion on reproducing functions) that we have a Lieb-Robinson bound for quasi-adiabatic evolution; that is,
if we evolve an operator $O$ under the equation of motion 
\be
\label{Odress}
\partial_s O(s)=i[{\cal D}_s,O],
\ee
we can prove a Lieb-Robinson
bound for $O(s)$.
This is a particular advantage compared to the Gaussian case, where the proof of Lieb-Robinson bounds for quasi-adiabatic
evolution is much more difficult and the bounds are weaker\cite{hall}.

Note that, as discussed previously in subsection (\ref{otherdecay}), we do not actually obtain a finite Lieb-Robinson
velocity.  That is, if we want to bound the commutator $\Vert [A_X(s),B_Y] \Vert$, where $X$ is supported on $X$ and $B$ is supported
on $Y$, the largest value of $s$ for which we obtain a meaningful bound grows slower than linearly in ${\rm dist}(X,Y)$.   This
is not a problem in most applications, since in general we will be considering path lengths of order unity, while we will
often consider distances between sets $X,Y$ which are of order system size.

\subsection{Goldstone's Theorem}
We now present an application to a non-relativistic Goldstone theorem.  This theorem is perhaps not that surprising, but the
results here (originally in \cite{goldstone}) are more general and simpler than previous nonrelativistic results\cite{ergodic}.

Above, we have described the clustering of correlation functions,
proving that the connected correlated function,
$\langle A B \rangle - \langle A P_0 B \rangle$,
of two operators $A,B$ with support on sets $X,Y$
is exponentially small in the distance ${\rm dist}(X,Y)$.  Here, for a system
with $k$ different ground states.
$\langle O \rangle \equiv k^{-1} \sum_{a}
\langle \psi_0^a, O \psi_0^a \rangle$.  For a parameter dependent
Hamiltonian, we define
$\langle O \rangle_s \equiv k^{-1} \sum_{a}
\langle \psi_0^a(s), O \psi_0^a(s) \rangle$.

The goal now is to prove (or at least sketch the proof of) a stronger statement about the decay of
correlation functions in gapped systems with a continuous symmetry showing
that the expectation value $\langle A P_0 B \rangle$ is small also.

Goldstone's  theorem is a statement that a system with a spontaneously broken continuous symmetry has gapless excitations.
We first need to define a continuous symmetry, or equivalently, a conserved charge.
This means that 
\begin{definition}
We say that a lattice Hamiltonian $H$ has a conserved charge if the following holds.
For every site $i$, there is an operator $q_i$ supported on site $i$,
with $q_i$ having integer eigenvalues.  Let $Q=\sum_i q_i$.  Then, we require that
\be
[Q,H]=0.
\ee
\end{definition}
Further, we assume that $\Vert q_i \Vert \leq q_{max}$ for some $q_{max}$ (this is a technical point, needed
in the later bounds).
Our non-relativistic Goldstone's theorem will be the contrapositive of the usual statement of Goldstone's theorem: we will
show that the presence of a gap (between a degenerate ground state sector and the rest of the spectrum) bounds the correlation
functions.

For any set $X$, we define $R(\theta,X)=\prod_{i\in X} \exp[i q_i \theta]$.
We consider operators $\phi_X,\overline \phi_Y$ with support on sets $X,Y$
which transforms as vectors as follows under this $U(1)$ symmetry:
$R(-\theta,X) \phi_X R(\theta,X)=\exp[i \theta] \phi_X$ and
$R(-\theta,Y) \overline \phi_Y R(\theta,Y)=\exp[-i \theta] \overline \phi_Y$.

For example, in a Bose system with conserved particle, the $q_i$ can represent
the particle number on a given site and the operators $\phi_X,\overline \phi_Y$
can represent creation and annihilation operators for the bosons.  For a
spin system, the $q_i$ can represent the $z$ component of the spin on a site
and the $\phi_X,\overline \phi_Y$ can represent raising and lowering spin
operators on sites.

We do not require the states in the ground state sector to be degenerate with
each other, simply the existence of a gap between that sector and the rest
of the spectrum.  This result is stronger than that in \cite{koma} as
it is valid in arbitrary dimension; it is also stronger than
other previous results\cite{ergodic} which either required a unique
ground state or else assumed an ergodic property which is equivalent
to requiring the vanishing of the matrix elements in the ground state
sector in which case the decay or correlations becomes equivalent to
clustering.

We will show that, for a local Hamiltonian on a finite dimensional lattice with a gap $\Delta E$
between the ground state
sector and the rest of the spectrum that
$\langle \phi_X \overline \phi_Y \rangle$ is superpolynomially small in ${\rm dist}(X,Y)$ (in \cite{goldstone}, stronger exponential
results are obtained).
To show this, we define a set of parameter dependent Hamiltonians 
${\cal H}_{\theta}$ as follows.
Let $X'$ denote the set of sites $i$
such that ${\rm dist}(X,i)\leq {\rm dist}(X,Y)/2$, as shown
in the figure.  Then define
$H_\theta=R(X',\theta) H R(X',-\theta)$.
Clearly, then, as $R(X',-\theta)$ is a unitary transformation,
${\cal H}_{\theta}$ has the same spectrum of ${\cal H}$ and
the ground states of ${\cal H}_{\theta}$ are given by
$\psi_0^a(\theta)\rangle=R(X',\theta) \psi_0(\theta)\rangle$.

\begin{figure}[tb]
\centerline{
\includegraphics[scale=0.7]{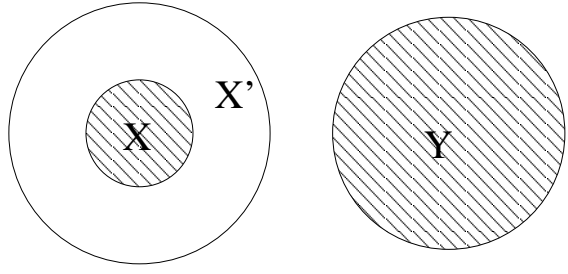}}
\caption{Illustration of the geometry we consider.  $X,Y$ are shown as shaded regions, while $X'$ includes everything within the outer circle around $X$.}
\end{figure}

Thus, 
\begin{eqnarray}
\label{cwt}
\partial_{\theta} \langle \phi_X \overline \phi_Y \rangle_{\theta}
\\ \nonumber
=\partial_{\theta} \langle R(X',-\theta) \phi_X \overline \phi_Y 
R(X',\theta) \rangle \\ \nonumber
=\partial_{\theta} \exp[i \theta] \langle \phi_X \overline \phi_Y 
\rangle \\ \nonumber
=i \langle \phi_X \overline \phi_Y 
\rangle,
\end{eqnarray}
where we used the fact that $X\subset X'$ while $Y\cap X'=0$ so that
$[\overline \phi_Y,R(X',\theta)]=0$ and where we evaluate the derivatives
at $\theta=0$.

Using Eq.~(\ref{P0change}), however,
\begin{eqnarray}
\label{cwd}
\partial_{\theta} \langle \phi_X \overline \phi_Y \rangle_{\theta}
\\ \nonumber
=
\frac{1}{q}
\partial_{\theta}
{\rm Tr}(P_0(\theta) \phi_X \overline \phi_Y )
\\ \nonumber
=
\frac{1}{q}{\rm Tr}(P_0(\theta) [\phi_X \overline \phi_Y,{\cal D}_\theta)]).
\end{eqnarray}
However, recall that ${\cal D}_{\theta}$ is a sum of terms ${\cal D}_Z(\theta)$, arising
from the different terms $\partial_theta H_Z(\theta)$.
We have $H_Z(\theta)=R(X',\theta) H_Z R(X',-\theta)$.
However, if $Z$ is a subset of $X'$ or if $Z$ is a subset of the complement of $X'$, then
$H_Z(\theta)=H_Z(0)$.  To see this, note that if $H_Z$ is a subset of the complement of $X'$,
then $H_Z$ commutes with $R(X',\theta)$.  If $H_Z$ is a subset of $X'$, then
$H_Z(\theta)=R(\Lambda,\theta) H_Z R(\Lambda,-\theta)$, where $\Lambda$ is the set of all sites.
Since $H_Z(\theta)$ commutes with $Q$, 
$R(\Lambda,\theta) H_Z R(\Lambda,-\theta)=H_Z$.
Thus, the only terms that contribute to ${\cal D}_\theta$ are indeed those where $Z$ intersects both
$X'$ and the complement of $X'$.  However, the corresponding terms ${\cal D}_Z(\theta)$ have
small commutator with $\phi_X \overline \phi_Y$ by the locality of the quasi-adiabatic evolution operator:
we can approximate ${\cal D}_\theta$ by an operator localized near the boundary of $X'$, a distance $l$ from
sets $X$ and $Y$ and so the commutator 
$[\phi_X \overline \phi_Y,{\cal D}_\theta)]$ can be shown to be superpolynomially small after summing over $Z$.

It is interesting to note that the assumption of a finite dimensional lattice is necessary in this
derivation (it comes in when we sum over $Z$, and is needed to bound the sum of terms by the number of terms times a bound on the norm
of each term).
We sketch a system which is not finite dimensional, and
show how a Goldstone theorem may fail in this case.
Consider a random graph with $V$ nodes each having coordination number
$3$.
Consider a set of $V$ spin-$1/2$ spins, with Hamiltonian
\be
H=-\sum_{i,j} J_{ij} \vec S_i \cdot \vec S_j,
\ee
where the interaction matrix $J_{ij}$ equals $1$ if the nodes $i,j$
are connected by an edge on the graph, and zero otherwise.  The interaction
is ferromagnetic, so pointing all spins up (or in any other direction)
gives a ground state.  Further, the Hamiltonian is local, using a shortest
path metric on the graph to define ${\rm dist}(i,j)$.
However, a random graph of this form is typically an expander 
graph\cite{expander} with a gap in the spectrum of the
graph Laplacian,
so a spin-wave theory calculation\cite{spinwave}
gives a gap in the magnon spectrum.  Thus, this system has a set of degenerate
ground states and a gap.  However, the spin correlations do not decay,
as $\langle \vec S_i \cdot \vec S_j \rangle=1/4$ for all $i,j$.

\section{Lieb-Schultz-Mattis in Higher Dimensions}
The Lieb-Schultz-Mattis theorem, proven in 1961\cite{lsm}, is a theorem about the spectrum on one-dimensional quantum spin systems
with symmetries.  We present the theorem in slightly more general form for theories with a conserved $U(1)$ charge, as considered later by Affleck and Lieb\cite{la}.

Consider a one-dimensional Hamiltonian $H$, with finite-range interactions (one can consider also sufficiently rapidly decaying
interactions; we do not consider this case in order to make the discussion as simple as possible but we encourage the reader
to work out what kinds of decay would still allow the theorem to be proven).  Assume that the Hamiltonian is translationally
invariant, with periodic boundary conditions.  Let $T$ be the translation operator, so $[T,H]=0$.

Then,
\begin{theorem}
Consider a one-dimensional, periodic, translationally invariant Hamiltonian with finite-range interactions, with $N$ sites, conserved charge $Q$ and ground state $\psi_0$.  Define the ground state
filling factor $\rho$ by
\be
\rho=\langle \psi_0, Q \psi_0 \rangle/N.
\ee
Assume that $\rho$ is not an integer.  Then, either the ground state is degenerate or the gap between the ground state and the
first excited state is bounded by
\be
\Delta E \leq {\rm const.}/N,
\ee
where the constant depends only on the strength $J$ of the interactions in $H$ and the range of the interactions.
\end{theorem}

Let us give some examples of the application of this theorem.  Consider the one-dimensional Heisenberg model:
\be
H=\sum_{i} \vec S_i \cdot \vec S_{i+1},
\ee
with spin-$1/2$ on each site.  We identify the charge $q_i$ by
\be
q_i=S^z_i+1/2,
\ee
so that $q_i$ has integer eigenvalues.  The Heisenberg model has not just the $U(1)$ invariance, but instead has a full
$SU(2)$ invariance (invariance under rotation).  So, if the ground state is non-degenerate, then the ground state has spin $0$.
Let us indeed assume that it is true that the ground state has spin $0$ (one can also prove this by other means), so also the ground state has $S^z=0$, which corresponds to $\rho=1/2$, hence $\rho$ is non-integer.  Thus, this model meets the conditions of the theorem and so must obey the conclusions: there must be a state within energy of order $1/N$ of the ground state.  In fact, the lowest
energy state has energy of order $1/N$ above the ground state, and corresponds to a ``spinon" excitation.  The model has a continuous energy spectrum in the infinite $N$ limit.

Another model is the Majumdar-Ghosh model\cite{mg}:
\be
H=\sum_{i} \vec S_i \cdot \vec S_{i+1}+(1/2)
H=\sum_{i} \vec S_i \cdot \vec S_{i+2}.
\ee
This model also meets the conditions of the theorem (it again has $SU(2)$ invariance and the ground state turns out to have total spin $0$).
So, it must meet the conclusions of the theorem.  However, this model meets the conclusions of the theorem in a different way.  It has
two exactly degenerate ground states and then a gap to the rest of the spectrum.  One ground state has spins $1$ and $2$ in a singlet, spins $3$ and $4$ in a singlet, and so on.  The other ground state is translated by one, so it has spins $2$ and $3$ in a singlet, and so on, and finally spins $N$ and $1$ in a singlet.  So, both ground states are products of singlets.
If the constant $1/2$ is changed to some number near $1/2$, then there appears an exponentially small splitting between the two
ground states, and a gap to the rest of the spectrum, which stil meets the conclusions of the theorem.

A useful exercise for the reader is the following: consider the state which is the symmetric combination of the two ground states
of the Majumdar-Ghosh model mentioned above, and call this state $\psi_0$.  It is an eigenvector of $T$ with eigenvalue $+1$.  Now, construct the state
$\psi_{LSM}$ and verify that it is close to (within distance $1/N$) the state which is the anti-symmetric combination of the
two ground states and that it is an eigenvector of $T$ with eigenvalue $-1$.

\begin{figure}
\centering
\includegraphics[width=200px]{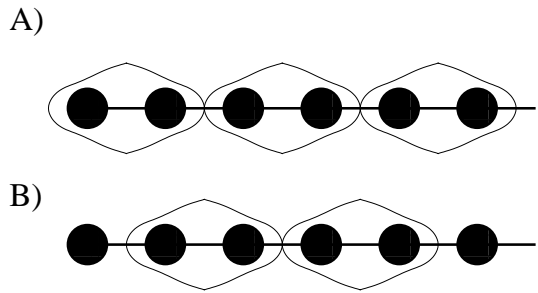}
\caption{A) One ground state of Majumdar-Ghosh model.  Circles indicate lattice sites.  Light line around circles indicate
that they are in a singlet.  B) Another ground state.}
\end{figure}

Finally, one can also consider spin systems with spin $1$ per site.  These systems do {\it not} meet the conditions of the theorem,
since if they have total $S^z=0$, then they have integer $\rho$.  Hence, the theorem does not imply the existence of a gap for
these systems.  In fact, such systems may have a unique ground state and a spectral gap, the so-called ``Haldane gap"\cite{haldane}.

We now prove the one dimensional Lieb-Schultz-Mattis theorem.
The proof is variational.  First, note that given that $[H,Q]=0$, without loss of generality we can assume that each term
in $H$ commutes with $Q$.  To prove this, let $H=\sum_Z H_Z$.  Then,
\be
H=\sum_Z H'_Z,
\ee
where
\be
H'_Z\equiv
\frac{1}{2\pi} \int_0^{2\pi} {\rm d}\theta \exp(i \theta  Q) H_Z \exp(-i \theta Q).
\ee
However, each term $H'_Z$ commutes with $Q$.  So, from now on, without loss of generality, we assume that every term in $H$ commutes
with $Q$.

Define a state $\psi_{LSM}$ by
\be
\psi_{LSM}=\Bigl( \prod_{j=1}^N \exp(2\pi i \frac{j}{N} q_j) \Bigr) \psi_0.
\ee
One may show that 
\be
\label{eq1}
\langle \psi_{LSM},H \psi_{LSM} \rangle - \langle \psi_0 , H \psi_0 \rangle \leq {\rm const.}/N.
\ee
To show this, we first show for every $Z$ that
\be
\label{eq2}
\langle \psi_{LSM},H_Z \psi_{LSM} \rangle - \langle \psi_0 , H_Z \psi_0 \rangle \leq {\rm const.}/N^2,
\ee
as may be proven using the fact that $Z$ has bounded diameter and that $H_Z$ commutes with $Q$ (we encourage the
reader to work through the detailed proof; the bound on the right-hand side will depend on the diameter of $Z$).
We then sum Eq.~(\ref{eq2}) over $Z$ to arrive at Eq.~(\ref{eq1}).

So, we have shown that $\psi_{LSM}$ is close in energy to the ground state.  We now show that $\psi_{LSM}$ is orthogonal to
the ground state, which will complete the variational proof of this one-dimensional Lieb-Schultz-Mattis theorem.  Note that we
can assume that the ground state is an eigenvector of $T$ (otherwise, since $T$ commutes with $H$, the ground state is degenerate), so that
\be
T \psi_0=z \psi_0,
\ee
for some complex number $z$ with $|z|=1$.  
Now consider $T\psi_{LSM}$.  We will show that $\psi_{LSM}$ is also an eigenvector of $T$ but with an eigenvalue different from $z$, which
will imply that $\psi_{LSM}$ is orthogonal to $\psi_0$, completing the proof.
We have:
\begin{eqnarray}
T\psi_{LSM} &=&
T \Bigl( \prod_{j=1}^N \exp(2\pi i \frac{j}{N} q_j) \Bigr) \psi_0 \\ \nonumber
&=& \Bigl\{ T \Bigl( \prod_{j=1}^N \exp(2\pi i \frac{j}{N} q_j) \Bigr) T^{-1}\Bigr\} T \psi_0 \\ \nonumber
&=& z \Bigl\{ T \Bigl( \prod_{j=1}^N \exp(2\pi i \frac{j}{N} q_j)\Bigr) T^{-1}\Bigr\}  \psi_0 \\ \nonumber
&=& z \Bigl( \prod_{j=1}^N \exp(2\pi i \frac{j}{N} q_{j-1}) \Bigr)  \psi_0,
\end{eqnarray}
where we define $q_0=q_N$ (recall that we have periodic boundary conditions).
Thus, shifting the summation variable $j$ by $1$ and recalling that $q_j$ has integer eigenvalues so $\exp(2 \pi i q_N)=1$, we have
\begin{eqnarray}
T\psi_{LSM} 
&=& z \Bigr( \prod_{j=1}^N \exp(2\pi i \frac{j+1}{N} q_j)  \Bigl)  \psi_0 \\ \nonumber
&=& z \Bigr( \prod_{j=1}^N \exp(2\pi i \frac{j}{N} q_j)  \Bigl) \exp(2\pi i Q/N) \psi_0 \\ \nonumber
&=& z \Bigr( \prod_{j=1}^N \exp(2\pi i \frac{j}{N} q_j)\Bigl)  \exp(2\pi i \rho) \psi_0 \\ \nonumber
&=& z \exp(2\pi i \rho) \psi_{LSM}.
\end{eqnarray}
In the above equation, we have assumed, without  loss of generality, that $\psi_0$ is an eigenvector of $Q$ (otherwise the ground
state degeneracy follows automatically since $[Q,H]=0$), so $\exp(2\pi i Q/N) \psi_0=\exp(2\pi i \rho) \psi_0$.
However, since we assume that $\rho$ is not an integer, we find that $\psi_{LSM}$ is an eigenvector $T$ with eigenvalue
$z \exp(2\pi i \rho) \psi_{LSM}$ which differs from $z$, completing the proof.

The above proof technique simply does not work in two dimensions.  Suppose we have a two-dimensional system, with size $L$ in each
direction which is periodic in one direction.  The energy of the state $\psi_{LSM}$ is a {\it constant} amount above the ground state, and
does not go to zero as $L$ goes to infinity.  The reason is that the energy per site is of order $1/L^2$, while there are a total of $L^2$
sites.  In contrast, in the one-dimensional case, there are only $L$ sites.  This different scaling between one and two dimensions should
be very familiar from statistical mechanics: two dimensions is the ``lower critical dimension" to break a continuous symmetry.

There is another physical reason why the one dimensional proof fails in two dimensions.  In one dimension, there are only
two possibilities\cite{la}.  Either, the system has a continuous spectrum, or, if it has a degenerate ground state and a spectral
gap (as in the case of the Majumdar-Ghosh model), there is a discrete symmetry breaking, with a local order parameter.  For example,
in the Majumdar-Ghosh model, for any $i$, the operator  $\vec S_i \cdot \vec S_{i+1}$ has non-trivial action in the ground state
subspace.  In contrast, in two dimensional system, there might also be topological order.  We might have a system with degenerate
ground states but for which no local operator has non-trivial action in the ground state subspace (i.e., every local operator is close
to a multiple of the identity when projected into the ground state subspace).

Topological order can arise in spin systems; one set of proposals involves the idea that a so-called ``short-range resonating valence bond" state describes the ground state\cite{rvb}.  Such states are a liquid-like superposition of various singlet configurations, that are in many
ways very physically similar to dimer models or to the toric code.  So, already on physical grounds we expect that
we will need some new technique to prove a Lieb-Schultz-Mattis theorem in higher dimensions.  Conversely, we get a nice payoff
from such a theorem in more than one-dimension: it will rule out the possibility (for certain systems which obey the conditions of
the theorem) of having a unique ground state and a spectral gap.  If we show then that some system does have a spectral gap then either
there is ordinary order (some local operator has nontrivial action in the ground state subspace) or there is topological order.
So, such a theorem can be a route to proving topological order.

We now sketch the higher dimensional proof\cite{hast-lsm,suffcon}.   This is intended only to be a sketch.  The statement is that:
\begin{theorem}
Consider a Hamiltonian $H$, defined on a finite-dimensional lattice with finite interaction range $R$ and a bound on interaction strength $J$.  Let $H$ have translation invariance in one direction with periodic boundary conditions and have a length $L$ in that direction.  Let the lattice have a total of $N$ sites, with $N$ bounded by
a constant times a polynomial in $L$.  Let $H$ have
conserved charge $Q$ and ground state $\psi_0$.  Define the ground state
filling factor $\rho$ by
\be
\rho=\langle \psi_0, Q \psi_0 \rangle/L.
\ee
Assume that $\rho$ is not an integer.  Then, either the ground state is degenerate or the gap between the ground state and the
first excited state is bounded by
\be
\Delta E \leq {\rm const.}\log(L)/L,
\ee
where the constant depends only on $R$,$J$, $q_{max}$, and the lattice geometry.
\end{theorem}
The theorem can be extended to sufficiently rapidly decaying interactions also.  The bound that $N$ is at most a polynomial times
$L$ implies that the theorem works for aspect ratios of order unity (for example, an $L$-by-$L$ square lattice in two dimensions) or even
aspect ratios which are quite far from unity (an $L$-by-$L^3$ square lattice, for example).  Note that the bound is slightly weaker
than in one dimension (we have $\log(L)/L$ instead of $1/L$).

Finally, the fact that
$\rho=\langle \psi_0, Q \psi_0 \rangle/L$ is a minor annoyance in the statement of the theorem.  In a spin-$1/2$ system on a square
latttice with odd width (i.e., an $L$-by-$M$ lattice with translational invariance in the first direction and with $M$ even) we do
indeed find that $\rho$ is non-integer.  However, for such a system on a lattice of even width, the theorem does not work.  In fact,
there are counterexamples to a conjectured theorem with even width (a spin ladder, consisting of an $L$-by-$2$ system of spin-$1/2$ spins
with Heisenberg $\vec S_i \cdot \vec S_j$ interactions between nearest neighbor spins has a unique ground state and a gap).  However,
we expect that if a two-dimensional system on an $L$-by-$M$ lattice has translation invariance and periodic boundary conditions in both directions, then the goes to zero as both $L$ and $M$ get large.  This has not been proven yet.  Still, the theorem above covers a wide variety of cases with a minimal number of assumptions (only one direction of translation invariance required).

Further, the translation invariance in at least one direction is a necessary condition.  The reader is invited to work out a counter-example
if no translation invariance is assumed.

The sketch of the proof is as follows.  It again is variational.  It is also a proof by contradiction.  That is, we assume that
the Hamiltonian has a spectral gap $\Delta E$ and use this assumed spectral gap to construct a variational state which has low energy.
If the initial gap is large enough (larger than a constant times $\log(L)/L$) the variational state will have energy less than
$\log(L)/L$ proving the theorem by contradiction.

We begin by defining a parameter-dependent family of
Hamiltonians, $H_\theta$.  These Hamiltonians are defined by ``twisting the boundary conditions" in one particular direction, the
direction in which the lattice is translation invariant.  Let us label the coordinate of a site $i$ in this direction by $x(i)$,
with $0\leq x(i)<L$.
To define the flux insertion operator, we need to define the Hamiltonian with twisted
boundary conditions.  Let $Q_X$ be defined by
\be
Q_X =\sum_{i}^{1\leq x(i)\leq L/2} q_i,
\ee
where $x(i)$ is the $\hat x$-coordinate of site $i$.  That is, $Q_X$ is the total charge in
the half of the system to the left of the vertical line with $x=L/2+1$ and to the right of $x=0$.
Let
\be
H(\theta_1,\theta_2)=\sum_Z H_Z(\theta_1,\theta_2),
\ee
where $H_Z(\theta_1,\theta_2)$ is defined as follows.
If the set $Z$ is within distance $R$ of
the vertical line $x=0$, then $H_Z(\theta_1,\theta_2)=\exp(i \theta_1 Q_X) H_Z \exp(-i \theta_1 Q_X)$;
if the set $Z$ is within distance $R$ of
the vertical line $x=L/2$, then $H_Z(\theta_1,\theta_2)=\exp(-i \theta_2 Q_X) H_Z \exp(i \theta_2 Q_X)$;
otherwise, $H_Z(\theta_1,\theta_2)=H_Z$.
Note that,
\be
\label{equivalent}
H(\theta,-\theta)=\exp(i \theta Q_X) H \exp(-i \theta Q_X).
\ee
This unitary equivalences implies that $H(\theta,-\theta)$ has the same spectrum as $H$ which will be useful below.

The introduction of the two different vertical lines is an important technical trick.  We now define an operator $W_1$ which generates the quasi-adiabatic evolution along the path where $\theta_1$ evolves from $0$ to $2\pi$ and $\theta_2=0$.  That is, define ${\cal D}^1_\theta$ to
generate the quasi-adiabatic evolution for $H_s=H(s,0)$ and let
\be
W_1=\exp(\int_0^{2\pi} d\theta {\cal D}^1_\theta),
\ee
where the exponential is $\theta$ ordered.  Similarly, let $W_2$ generate quasi-adiabatic evolution along the path where $\theta_2$ evolves
from $0$ to $2\pi$ and $\theta_1=0$ and let ${\cal D}^2_\theta$ generate the quasi-adiabatic evolution for $H_s=H(0,-s)$.
Finally, let $W$ generate quasi-adiabatic evolution along the path $\theta_1=-\theta_2=\theta$ as $\theta$ evolves from $0$ to $2\pi$.

An important point: we do {\it not} assume that the gap remains open along the paths above used to define $W_1,W_2$.  We simply use the
assumed initial gap at $\theta_1=\theta_2=0$ and then evolve quasi-adiabatically as if the gap remained open along the path.

Using locality of the quasi-adiabatic evolution operators, if the gap is sufficiently large,
one may show that
\be
W_1 W_2 \approx W_2 W_1 \approx W,
\ee
where the approximation means that $W_1 W_2-W$ is small in operator norm.  The size that the gap needs to be and the magnitude of
the error $\Vert W_1 W_2-W \Vert$ both depend on the quasi-adiabatic evolution operator we use.  Roughly, the quasi-adiabatic evolution
operator ${\cal D}^1$ is supported near the line $x(i)=0$, up to a length scale which is inversely proportional to the gap.  We need this length scale to be small compared to $L/4$ so that we can approximate the operator ${\cal D}^1$ by an operator supported within distance $L/4$ of
the line $x(i)=0$; we make the same approximation for ${\cal D}^2$ so that in this case, the operators ${\cal D}^1$ and ${\cal D}^2$ can be approximated by operators supported on disjoint
sets.
Using exact quasi-adiabatic evolution operators, one finds that the gap needs to be at least $f(l)/L$ for some function $f$ growing
slower than any polynomial, while for Gaussian operators one can choose $f(L)$ to be a constant times $\log(L)$.

\begin{figure}
\centering
\includegraphics[width=200px]{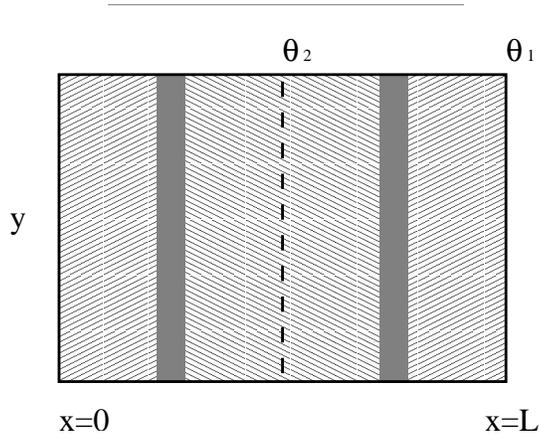}
\caption{Twist in boundary conditions are applied at two places, at $x(i)=0=L$ (along the boundary of the systems) and at $x(i)=L/2$ (along the dashed line).
We want to approximate $W_1$ by an operator supported on the upward slanting grey lines (near $x(i)=0$) and to approximate the operator
$W_2$ by an operator supported on the downward slanting grey lines, so that $W_1,W_2$ will approximately commute.  This requires a gap
sufficiently large compared to $1/L$.}
\end{figure}

We consider the variational state:
\be
W_1 \psi_0.
\ee
We wish to show that this state has low energy.  Note that $W_1$ is unitary.  The Hamiltonian $H$ is a sum of terms $H_Z$.  We will
show that each term $H_Z$ has roughly the same expectation value in the state
$W_1 \psi_0$ as it does in the ground state.  If $Z$ is far from the line $x(i)=0$ (for example, far can mean that
the distance is at least $L/4$), then this follows directly from the locality
of the quasi-adiatic evolution operator: $H_Z$ almost commutes with $W_1$:
\be
\langle W_1 \psi_0, H_Z W_1 \psi_0 \rangle \approx \langle \psi_0, H_Z \psi_0 \rangle.
\ee
Conversely, if $H_Z$ is near the lnie $x(i)=0$ (say, the distance is less than $L/4$), $H_Z$ almost commutes with $W_2$.  So:
\begin{eqnarray}
\langle W_1 \psi_0, H_Z W_1 \psi_0 \rangle & \approx & \langle \psi_0, W_1^\dagger W_2^\dagger H_Z W_2 W_1 \psi_0 \rangle \\ \nonumber
& \approx & \langle \psi_0, W^\dagger H_Z W \psi_0 \rangle.
\end{eqnarray}
Note, however, that $W$ describes evolution along a path of Hamiltonians which are all unitarily equivalent to $H(0,0)$ so all
Hamiltonians along this path have the same spectral gap.  Further, $H(2\pi,-2\pi)=H(0,0)$ So, $W \psi_0 =w \psi_0$ for some
complex number $w$ with $|w|=1$ (we will worry about this phase $w$ in the next paragraph, for now it doesn't matter).  So,
$\langle W_1 \psi_0, H_Z W_1 \psi_0 \rangle \approx \langle \psi_0, H_Z \psi_0 \rangle$ for all $Z$.

This proves that $W_1 \psi_0$ is a low energy state.  We now need to show that $W_1 \psi_0$ is orthogonal to the ground state.
We do this by showing that it has a different expectation value for the translation operator, $T$, than the ground state does.
Suppose
$\langle \psi_0,T \psi_0 \rangle=z$ for some complex number $z$.  Then,
consider $\langle \psi_0, W_1^\dagger T W_1 \psi_0 \rangle$.  This equals
$\langle \psi_0, W_1^\dagger (T W_1 T^{-1}) T\psi_0 \rangle=
z\langle \psi_0, W_1^\dagger (T W_1 T^{-1})\psi_0 \rangle$.
This is approximately equal to
$z\langle \psi_0, W_2^\dagger W_1^\dagger (T W_1 T^{-1})W_2 \psi_0 \rangle$.
Note that $W_2^\dagger W_1^\dagger$ is close to $W^\dagger$, so $W_1 W_2 \psi_0$ is close to $w \psi_0$ where $w$ is some
phase as mentioned above.  Similarly, 
$W_2^\dagger W_1^\dagger (T W_1 T^{-1})W_2 \psi_0$ is close to $w' \psi_0$ for some other complex number $w'$ with $|w'|=1$.
To see this, note that
$(T W_1 T^{-1})$ describes quasi-adiabatic evolution where we twist the boundary conditions along the line $x(i)=1$ rather than
along $x(i)=0$, so $(T W_1 T^{-1}) W_2$ is close to some operator $W'$ which describes quasi-adiabatic evolution of a Hamiltonian with
twisted boundary condition by $\theta$ along line $x(i)=1$ and by $-\theta$ along $x(i)=L/2$.
So, 
$\langle W_1 \psi_0,T W_1 \psi_0 \rangle$ is close to $\overline w w' z$.  We now just need to work out the phases $w,w'$.

However, using the property (\ref{Berryprop}), one can show that
$w=\exp(2\pi i \langle \psi_0, Q_X \psi_0 \rangle)$.  Similarly, the phase $w'$ depends on the expectation value of
$q_i$ summed over $1<x(i)\leq L/2$.  So,
\be
\overline w' w=\exp(2 \pi i \langle \psi_0, \sum_{i,x(i)=1} q_i \psi_0 \rangle).
\ee
This expectation value is non-integer by assumption.  So, $W_1 \psi_0$ has a different expectation value for $T$ than $\psi_0$,
so it is orthogonal to $\psi_0$.

Some heuristic comments on why physically our proof of the theorem is necessarily a proof by contradiction.  That is, why we assumed a gap
at the beginning of the proof.  We use the idea of twisting the boundary conditions.  A state such as an anti-ferromagnet will
strongly resist this twist in boundary conditions; that is, the ground state energy will change by an amount of order unity when we impose
this boundary  twist in a two-dimensional system (and by an even larger amount in higher dimensions).  Thus, for a state such as an anti-ferromagnetic (which is gapless), there is no reason to expect that the procedure we described of twisting boundary conditions and following  the quasi-adiabatic evolution of the state along the path will give any useful results.  However, suppose we have a system which has a gap.  By the theorem we have just sketched, such a system cannot have a unique  ground state and then a gap to the next lowest energy state.  So, we instead want to consider a state with a degenerate ground state and a gap.  Such a system could be a valence bond solid or a resonating valence bond
system, among other possiblities.  In such a system, twisting the boundary conditions does not lead to a large energy cost.  Instead, it
typically costs only an exponentially small amount of energy to twist the boundary conditions.  Then, when we start at one ground state and quasi-adiabatically evolve it, twisting the boundary angle $\theta$ from $0$ to $2\pi$, we transform it a state close to an orthogonal ground state.  This is analogous to the discussion in one dimension, where starting with one of the ground state of the Majumdar-Ghosh model and constructing the state $\psi_{LSM}$, gave us something close to the other ground state.  So, the physical idea of the higher-dimensional proof is that if there is a sufficiently large gap from the ground state sector to the rest of the spectrum, then we can use the quasi-adiabatic continuation to construct
a unitary that transforms one ground state into another.

\section{What is a Phase?}
What is a phase of a quantum many-body system?  We are used to the idea that physical systems have distinct phases, with phase
transitions between them.  For example, water can appear as ice, water, or steam (and further, there are many distinct phase of ice).
This discussion of the properties of water is a discussion of systems at non-zero temperature, while our focus in this notes is
on quantum systems at zero temperature, but many of the same phenomena occur in both cases.  For example, we do not actually consider
steam and water to be distinct phases of matter.  While usually water turns into steam by being boiled (a phase transition, where
the energy is non-analytic in the thermodynamic limit), we can also move from water to steam without any phase transition, by following
a path in the two-dimensional plane of temperature and pressure.  A similar phenomenon occurs in the transverse field Ising model.
Suppose we consider the model with an additional parallel magnetic field so that the Hamiltonian is
\be
\label{tfi2}
H=-J \sum_{i=1}^{N-1} S^z_i S^z_{i+1} + B \sum_{i=1}^{N} S^x_i+H \sum_{i=1}^N S^z_i.
\ee
Suppose $J>>B$ and $H>0$.  Then, the system has a unique ground state; at $B=0$, this ground state is the state with all spins
up, while for $B>0$, there are quantum fluctuations about this state.

When $H$ changes sign, the ground state changes from all
spins up to all spins down in the case $B=0$, crossing a phase transition.  This is a zero temperature phase transition.  In the
case of $B=0$, this phase transition is a level crossing: at $H=0$, there are two exactly degenerate ground states.
For $B\neq 0$, this level crossing becomes an avoided crossing which we now describe.  Recall that we said that for $B\neq 0$ but $H=0$, the system
has two ground states with an exponentially small splitting between them.  The behavior of the two lowest energy states
as a function of $H$ can be roughly understood in the following toy model.  Suppose that $H$ is very small.  Then we can
focus on just the two lowest energy states and use Eq.~(\ref{ordparam}) to arrive at the following two-by-two Hamiltonian
(we obtain this Hamiltonian by projecting the term
$\sum_{i=1}^N S^z_i$ into the ground state subspace):
\be
H=\begin{pmatrix}
t & HmN \\ HmN & -t,
\end{pmatrix},
\ee
where $t$ is some exponentially small splitting between the two lowest states and $N=|\Lambda|$ is the size of the system.

By a change of basis (going to symmetric and anti-symmetric combinations of the two ground states), we arrive instead at
\be
H=\begin{pmatrix}
HmN &t \\ t & -HmN
\end{pmatrix}.
\ee
The two different basis vectors here correspond to the spin up and spin down ground states.

\begin{figure}
\centering
\includegraphics[width=200px]{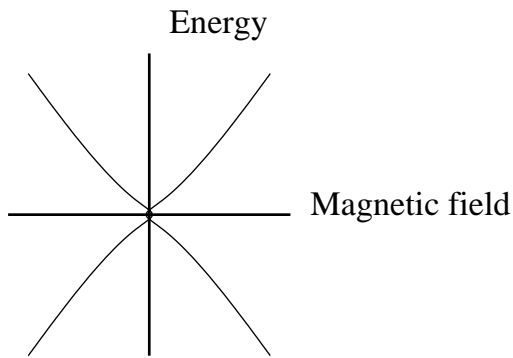}
\caption{Energy of lowest two states as a function of parallel magnetic field $H$.  This is a sketch.  The crossing is an avoided crossing but
the splitting between states is exponentially small at $H=0$.  At $H$ of order $1/N$, there is another avoided crossing as the energy gap becomes of order unity.}
\end{figure}

While this Hamiltonian is valid for small $H$, for larger $H$ we need to worry about the excited states.  However, this Hamiltonian
already reveals the essential point, namely that since $t$ is exponentially small as a function of $N$, in the limit of $N\rightarrow \infty$,
the ground states energy per site is a non-analytic function of $H$.  However, this behavior is a lot like changing from water to steam by boiling: we can also move from $H<0$ to $H>0$ without crossing a phase transition.  Instead, one should follow the path of first making $H$ large
and negative, then decreasing $J$ until $J<<B$, then changing the sign of $H$ (which does not involve a phase transition since $J<<B$) and then increasing $J$, as shown in Fig.~(\ref{pathfig}).

\begin{figure}
\centering
\label{pathfig}
\includegraphics[width=200px]{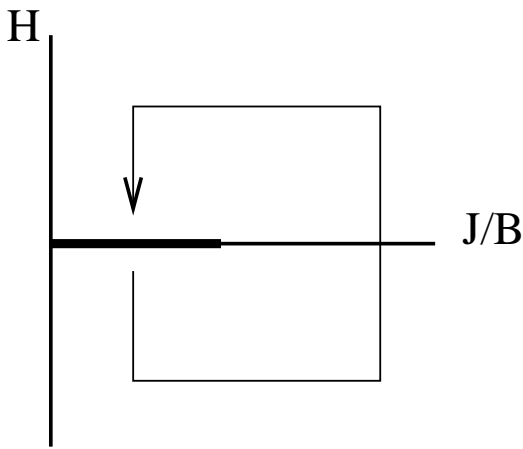}
\caption{Path to follow.  Thickened line on axis denotes $B/J$ less than the critical value.}
\end{figure}

We are motivated by this  analysis to adopt the following definition of a quantum phase: two Hamiltonians, $H_0$ and $H_1$ describe
systems in the same quantum phase if both $H_0$ and $H_1$ have a spectral gap, and one can find a smooth path $H_s$ connecting $H_0$
and $H_1$ which keeps the gaps open and keeps the Hamiltonian local.
  (As a technical point for those interested, in some cases
it may be more appropriate to consider a ``stable limit" when describing equality of quantum phases, as in the case of topologically
ordered phases of free fermion systems\cite{kitaevtable}).

Now, one may choose to have a more refined notion of a quantum phase, which
takes into account symmetries; for example, one might wish to insist that there is a path connecting $H_0$ to $H_1$ which respects symmetries
such as an Ising symmetry.  We do not consider this kind of restriction on the definition here.  Our interest is instead the case of
systems which are in distinct phases even without any assumptions on the symmetry.  That is, Hamiltonians which {\it cannot} be connected
by such a smooth path.

We can use the ideas developed in these notes to prove that certain Hamiltonians cannot be connected by a smooth path of local Hamiltonians
without closing the gap or without the path length being long (where ``long" means that the path length diverges as the system size gets large).  Consider, for example, a toric code on a torus.  This has four ground states and a spectral gap.  Call this Hamiltonian $H_0$.  Consider instead a Hamiltonian consisting of two copies of the transverse field Ising model on a torus with $J>>B$.  Call this Hamiltonian $H_1$ (if one wants a better statement of $H_1$, break the square lattice into two different sublattices, and have interactions only between spins on a given sublattice, so that way we have two copies of the transverse field Ising model with the same number of degrees of freedom as in the toric code system).  Note that $H_0,H_1$ both have $4$ ground states and a gap.  So, can we find a path connecting $H_0$ to $H_1$?
The answer is no.  If such a path existed, then we could use quasi-adiabatic continuation to evolve the four ground states of $H_0$ to
produce some linear combination of the four ground state of $H_1$.  However, this would imply that (recall the discussion in subsection \ref{toute} and the fact that we have a Lieb-Robinson bound for the quasi-adiabatic continuation operators) the ground states of $H_1$ are also topologically ordered.  Since this is not true, no such path can exist.  The reason we need to assume that the path is not ``long" is that
the length of the path plays the role of time in the Lieb-Robinson bound, and recall that in the discussion of the behavior of topological
order under time evolution we only showed that topological order could not appear after evolution for a short time, but not for arbitrary time.

We can use similar arguments to show that Hamiltonians cannot be connected by such a smooth path even when the ground state is unique (again, without closing the gap and without the path being long).  One
of the key properties of systems like the toric code is the particular set of expectation values they have for certain
operators they have called string operators; these are operators which are products of
single-site operators around a loop which are analogous to Wilson loops in gauge theories.  One can define ``dressed operators" by quasi-adiabatically continuing these string operators along a path.  Thus, given a string operator $O$ for a Hamiltonian $H_0$,
we can define an operator
\be
\tilde O \equiv U O U^\dagger,
\ee
where
\be
U={\cal S} \, \exp(i\int_0^1 {\rm d}s {\cal D}_s),
\ee
where the calligraphic $s$ in front of the exponential denotes that it is an $s$-ordered exponential.  Then, the operator $\tilde O$ has the same expectation value in the ground
state of $H_1$ as $O$ does in the ground state of $H_0$.  
This dressed operator $O$ is precisely equal to the operator $O(s=1)$ as described by the evolution of Eq.~(\ref{Odress}).

Further, if two operators $O,O'$ anti-commute with each other, then the
operators $\tilde O,\tilde O'$ also anti-commute with each other.  Now, consider the transverse field Ising model in the phase
$B>>J$.  This has a unique ground  state.  Similarly, the toric code on a sphere has a unique ground state.  However, these
two models cannot be connected by a continuous path.  To show this, note that if they were connected, then we could also connect
the toris code to the transverse field Ising model in the phase $B\neq 0, J=0$.  However, the ground state of this model with $B\neq 0,J=0$
can be shown to be inconsistent with the properties of such dressed string operators.
Here is a sketch (this sketch depends on properties of the toric code which are not discussed in these notes  and need to be read elsewhere):
Consider an electric loop operator, indicated as the solid line in (A) of the figure.  Call this operator $E$.  Let the dashed line represent a magnetic loop operator, which we call $M$.
These operators commute with each and both have expectation value $1$ in the toric code ground state.  In (B) of the figure, we show that
the operator $E$ can be written as a product of two different operators, which we call $E_1$ and $E_2$ which act on part of the loop.
Note that $\{E_1,M\}=\{E_2,M\}=0$.  Hence, the expectation value of $E_1 M E_2$ is equal to minus 1 in the toric code ground state.  Now
consider the dressed operators $\tilde E_1,\tilde E_2,\tilde M$ assuming the existence of a path connecting the
toric code to the transverse field Ising model.  Since $\tilde E_1 \tilde E_2$ would have expectation value 1 in the transverse
field Ising model ground state, $\psi_1$, the state $\tilde E_1 \tilde E_2 \psi_1$ must be simply the product state of all spins pointing
along the transverse magnetic field.  However, using locality of the dressed operators, $\tilde E_1$ can be approximated by an operator
supported near the support of $E_1$.  Hence, the state $\tilde E_2 \psi_1$ must have all of the spins
which are far from the support of $E_1$ pointing approximately along the transverse magnetic field.  Thus, the state
$\tilde E_2 \psi_1$ has all of its spins, except those near the upper and lower ends of the support of $E_2$ (those close to the support of
both $E_1$ and $E_2$) pointing approximately along the magnetic field.
Similarly, the operator
$\tilde M$ can be approximated by an operator supported near the support of $M$; however, this means that $\tilde M$ can be approximated
by an operator, which  we call $\tilde M'$, which is not supported near the upper and lower ends of $E_2$.  As noted, the spins away from the upper and lower ends
of the support of $E_2$ are aligned approximately along the magnetic field, and those are the only spins in the support of $\tilde M'$.  Since
$\tilde M \psi_1=\psi_1$, we have $\tilde M' \psi_1$ close to $\psi_1$ and
so, using the fact that the spins in the support of $\tilde M'$ are almost aligned with the field in
the state $\tilde E_2 \psi_1$ and in the state $\psi_1$, we find that $\tilde M' \tilde E_2 \psi_1$ is close to $\tilde E_2 \psi_1$.  Thus, $\tilde M \tilde E_2 \psi_1$ is close to
$\tilde E_1 \psi_1$.
Hence, we find that the expectation value of $\tilde E_1 \tilde M \tilde E_2$ is close to unity in the transverse field Ising model,
while it was close to  minus 1 in the toric code, giving a contradiction.

\begin{figure}
\centering
\includegraphics[width=200px]{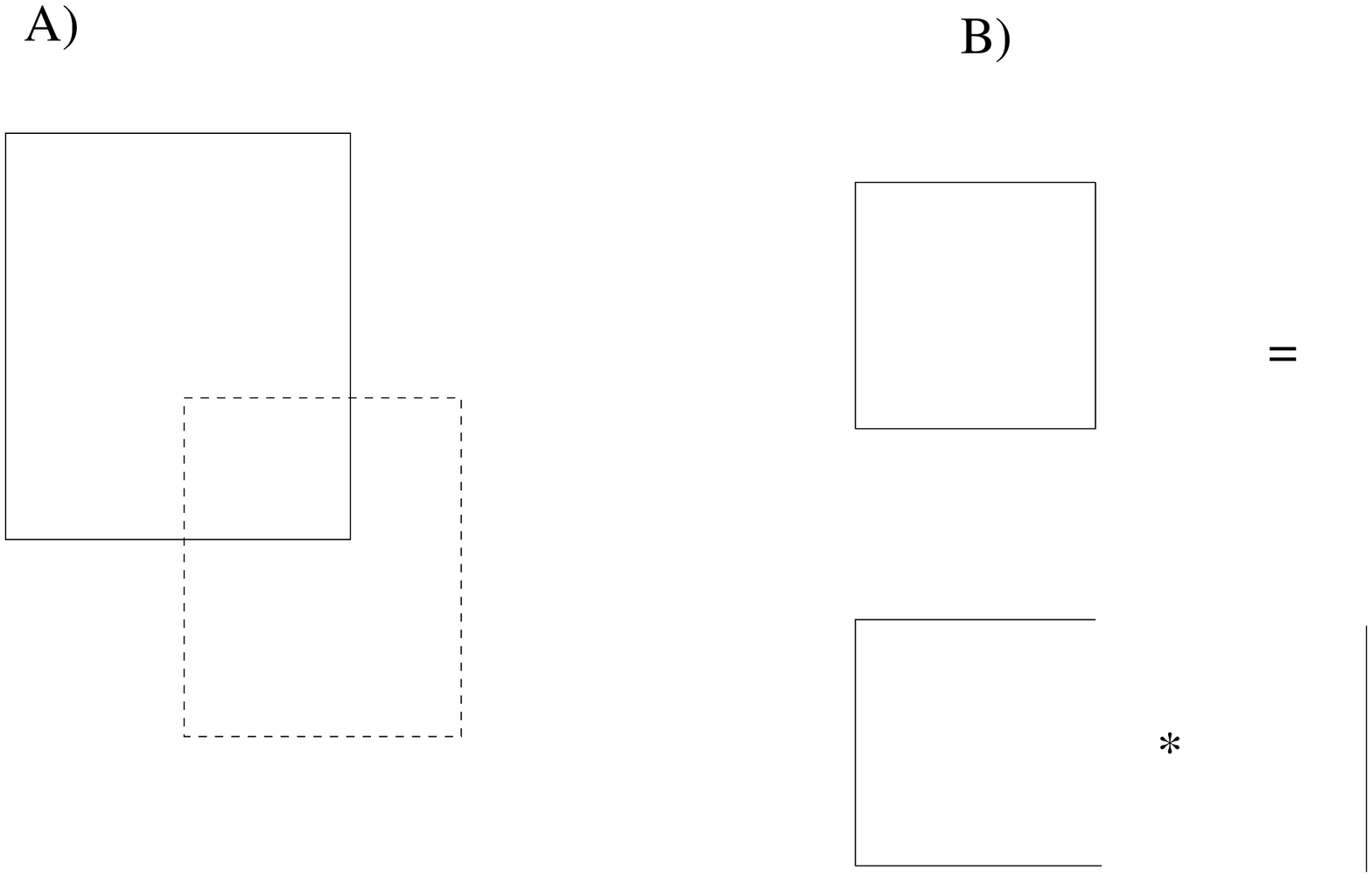}
\end{figure}

One may also choose to take into account other (anti-unitary) symmetries such as time-reversal symmetry.  These symmetries
are well-understood in the non-interacting case\cite{kitaevtable}, but only a limited understanding has been obtained in the
interacting case\cite{kitaev2}.  In general, even without symmetries the classification of phases of matter of interacting systems under the definition above
is only in the earliest stages.  One interesting case is that lattice Hamiltonians are known, the so-called ``Levin-Wen models"\cite{levinwenmodel}, which realize certain two-dimensional unitary topological quantum field theories (TQFTs), in particular those theories which are quantum doubles.  
Thus, a classification of interacting phases of matter requires a classification of these TQFTs.  Some results on classification of TQFTs were
obtained in \cite{rowell}.  However, it is likely that the full classification of lattice models includes many other phases in addition to
those described by TQFTs, so this classification is a problem for the future.

\section{Stability of Topologically Ordered Phase}
In this section, we briefly mention certain recent results on the stability of quantum phases.  The analysis throughout these
notes has always dealt with systems with a gap.  In some cases, we were able to prove either the absence of a gap or the degeneracy
of a ground states (such as in the Lieb-Schultz-Mattis theorem).  However, the reader may be wondering: how do we know a system
has a gap?  Similarly, suppose we have proven that a certain model, such as a toric code or Levin-Wen model, is in a topologically
nontrivial phase and cannot be connected to a topologically trivial phase without closing the gap.  However, what happens if we slightly
perturb the Hamiltonian?  Suppose we consider a Hamiltonian
\be
H=H_0 + sV,
\ee
where $H_0$ is some unperturbed Hamiltonian describing a topologically nontrivial phase, $s$ is some real number, and $V$ is a perturbation.
Does the model remain in the same phase for sufficiently small $s$?  Does the gap remain open?

The interesting question here is to consider the case in which $V$ is a sum of local terms.  Thus,
we want
\be
V=\sum_Z V_Z,
\ee
where the operator norms $\Vert V_Z \Vert$ decay rapidly as a function of the diameter of the set $Z$ (just as we required a similar
decay on the norms of the terms $\Vert H_Z \Vert$ in the Hamiltonian $H_0$).  A very elementary result is that for any given system size,
there is an $s_0$ such that for $|s|<s_0$ the gap remains open: simply use the fact that for any given system size, the norm of
the operator $V$ is finite, and the gap for $s>0$ is lower bounded by
\be
\Delta E(s) \geq \Delta E(0)- 2 s \Vert V \Vert,
\ee
where $\Delta E(s)$ denotes the gap as a function of $s$ and the factor of $2$ in front of the second term occurs because the ground
state energy increases by at most $s\Vert V \Vert$ while the first excited state energy decreases by at most the same amount.

So, we may take $s_0=\Delta E(0)/4\Vert V \Vert$.
However, such a bound, while elementary, is also fairly useless, since it leads to an $s_0$ which tends to zero as the system size
tends to infinity.  Instead, we want a bound which is a uniform function of system size.  Such bounds were provided in \cite{stability1,stability2}.  We will not review them here, except to note that using such bounds one can then prove (using quasi-adiabatic continuation) that
many of the properties of the topologically ordered system (such as ground state splitting, braiding, fusion rules, etc...) remain the
same in this phase.  An interesting open question is to understand the behavior of topological entanglement entropy as a function of perturbation.  Perhaps some smoothed definition of topological entanglement entropy exists (smoothing over different boundaries?) such that it can
also be proven to be invariant under perturbations?  These problems, and problems like the classification of different phase of lattice
quantum systems, are problems for the future.

{\it Acknowledgment} I thank S. Bravyi, T. Koma, T. Loring, S. Michalakis, F. Verstraete, and X.-G. Wen for useful collaboration on these and related ideas.


\begin{thebibliography}{99}
\bibitem{lr1} E. H. Lieb and D. W. Robinson, Commun. Math. Phys. {\bf 28},
251 (1972).

\bibitem{hast-lsm} M. B. Hastings, Phys. Rev. B {\bf 69}, 104431 (2004).


\bibitem{koma} M.  B. Hastings and T. Koma,
Commun. Math. Phys. {\bf 265}, 781 (2006).

\bibitem{ns} B. Nachtergaele and R. Sims, Commun. Math. Phys.
{\bf 265}, 119 (2006).

\bibitem{bhv} S. Bravyi, M. B. Hastings, and F. Verstraete,
``Lieb-Robinson Bounds and the Generation of Correlations
and Topological Quantum Order",
Phys. Rev. Lett. {\bf 97}, 050401 (2006).


\bibitem{mob} M. B. Hastings, arXiv:1001.5280.

\bibitem{locality} M. B. Hastings,
``Locality in Quantum and Markov Dynamics on Lattices and Networks",
Phys. Rev. Lett. {\bf 93}, 140402 (2004).

\bibitem{toric}  A. Kitaev,
{\it ``Fault-tolerant quantum computation by anyons"},
 Ann.~Phys.~{\bf 303}, 2 (2003).

\bibitem{levinwen} M. Levin and X.-G. Wen, arXiv:cond-mat/0510613,
Phys. Rev. Lett. {\bf 96}, 110405 (2006).

\bibitem{kitaevpreskill} A. Kitaev and J. Preskill, hep-th/0510092, 
Phys. Rev. Lett. {\bf 96}, 110404 (2006).

\bibitem{stability1}  S. Bravyi, M. B. Hastings, and S. Michalakis, {\it ``Topological quantum order: stability under local perturbations"},
e-print arXiv:1001:0344 (2010).


\bibitem{stability2}  S. Bravyi and M. B. Hastings, {\it ``A Short Proof of Stability of Topological Order Under Local Perturbations"},
e-print arXiv:1001:4363 (2010).


\bibitem{wen} X. G. Wen  and Q. Niu,
{\it ``Ground-state degeneracy of the fractional quantum Hall states in the presence of a random potential and on high-genus Riemann surfaces"},
Phys.~Rev.~{\bf B41}, p.~9377 (1990).


\bibitem{freedman} M. H. Freedman, A. Kitaev, M. J. Larsen, and Z. Wang,
{\it ``Topological Quantum Computation"},
e-print quant-ph/0101025 (2001).


\bibitem{note} A. E. Ingham, ``A Note on Fourier Transforms",
J. London Math. Soc. {\bf 9}, 29 (1934).


\bibitem{subexp} J. Dziubanski and E. Hern\'{a}ndez, ``Band-limited wavelets with subexponential decay",
{\bf 41}, 398 (1998).

\bibitem{hall} M. B. Hastings and S. Michalakis, ``Quantization of Hall Conductance for Interacting Electrons without Averaging Assumptions",
arXiv:0911.4706, Commun. Math. Phys., submitted.


\bibitem{goldstone} M. B. Hastings, ``Quasi-Adiabatic Continuation in Gapped Spin and Fermion
Systems: Goldstone's Theorem and Flux Periodicity", preprint
cond-mat/0612538, JSTAT, P05010 (2007).


\bibitem{osborne}  T. J. Osborne, ``Simulating adiabatic evolution of gapped spin systems",
Phys. Rev. A, {\bf 75}, 032321.


\bibitem{ergodic}   W. F. Wreszinski,
Fortschr. Phys. {\bf 35}, 379 (1987);
L. Landau, J. Fernando Perez and W. F. Wreszinski,
J. Stat. Phys. {\bf 26}, 755 (1981).


\bibitem{expander}  N. Alon, Combinatorica, {\bf 6(2)}, 83 (1986);
J. Friedman, Conf. Proc. of the Annual ACM
Symposium on Theory of Computing, 720 (2003); J. Friedman, preprint cs/0405020.


\bibitem{spinwave} N. W. Ashcroft and N. D. Mermin, {\it Solid
State Physics}, Chapter 33, (Harcourt Brace College Publishers, New York, 1976).

\bibitem{lsm}  E. H. Lieb, T. D. Schultz, and D. C. Mattis, Ann. Phys. (N. Y.)
{\bf 16}, 407 (1961).


\bibitem{la} I.Affleck and E.H.Lieb, Lett.Math.Phys. 12, 57 (1986).


\bibitem{mg} C. K. Majumdar and D. Ghosh, J. Math. Phys. 10, 1388 (1969).

\bibitem{haldane} F. D. M. Haldane, Phys. Rev. Lett. {\bf 50}, 1153 (1983).


\bibitem{suffcon} M. B. Hastings, Europhys. Lett. {\bf 70}, 824 (2005).

\bibitem{rvb} B. Sutherland, Phys. Rev. B {\bf 37}, 3786 (1988);
D. Rokhsar and S. Kivelson, Phys. Rev. Lett. {\bf 61}(1988);
N. Read and B. Chakraborty, Phys Rev. B {\bf 40}, 7133 (1989).

\bibitem{kitaevtable}  A. Kitaev, arXiv:0901.2686.

\bibitem{kitaev2} L. Fidkowski and A. Kitaev,  arXiv:0904.2197.

\bibitem{levinwenmodel}  M. A. Levin and X.-G. Wen,
{\it ``String-net condensation: A physical mechanism for topological phases"},
Phys.~Rev.~{\bf B71}, 045110 (2005).


\bibitem{rowell} E. Rowell, R. Stong, and Z. Wang, ``On classification of modular tensor categories", Comm. Math. Phys. {\bf 292}, 343 (2009), arxiv: 0712.1377.
\end{thebibliography}
\end{document}